# Commissioning Progress of the FAST

Peng Jiang[1*], Youling Yue[1], Hengqian Gan[1], Rui Yao[1*], Hui Li[1], Gaofeng Pan[1], Jinghai Sun[1], Dongjun Yu[1], Hongfei Liu[1], Ningyu Tang[1], Lei Qian[1], Jiguang Lu[1], Jun Yan[1], Bo Peng[1*], Shuxin Zhang[1], Qiming Wang[1], Qi Li[1], Di Li[1, 2], and FAST Project

[1] *CAS Key Laboratory of FAST, National Astronomical Observatories, Chinese Academy of Sciences, Beijing 100101, China.*

[2] *School of Astronomy and Space Science, University of Chinese Academy of Sciences, Beijing, 100049, China.*

*Correspondence should be addressed to Peng Jiang (pjiang@nao.cas.cn), Bo Peng (pb@nao.cas.cn) and Rui Yao(ryao@nao.cas.cn).*

Abstract: The Five-hundred-meter Aperture Spherical radio Telescope (FAST) was completed with its main structure installed on September 25, 2016, after which it entered the commissioning phase. This paper aims to introduce the commissioning progress of the FAST over the past two years. To improve its operational reliability and ensure effective observation time, FAST has been equipped with a real-time information system for the active reflector system and hierarchical commissioning scheme for the feed support system, which ultimately achieves safe operation of the two systems. For meeting the high-performance indices, a high-precision measurement system was set up based on the effective control methods that were implemented for the active reflector system and feed support system. Since the commissioning of the FAST, a low-frequency ultra-wideband receiver and 19-beam 1.05-1.45 GHz receiver have been mainly used. Telescope efficiency, pointing accuracy, and system noise temperature were completely tested and ultimately achieved the acceptance indices of the telescope. The FAST has been in the process of national acceptance preparations and has begun to search for pulsars. In the future, it will still strive to improve its capabilities and expand its application prospects.

## 1. Introduction

The Five-hundred-meter Aperture Spherical radio Telescope (FAST) is a Chinese mega-science project aimed to build the largest single dish and most sensitive radio telescope in the world, as shown in Fig. 1.1 [1, 2].

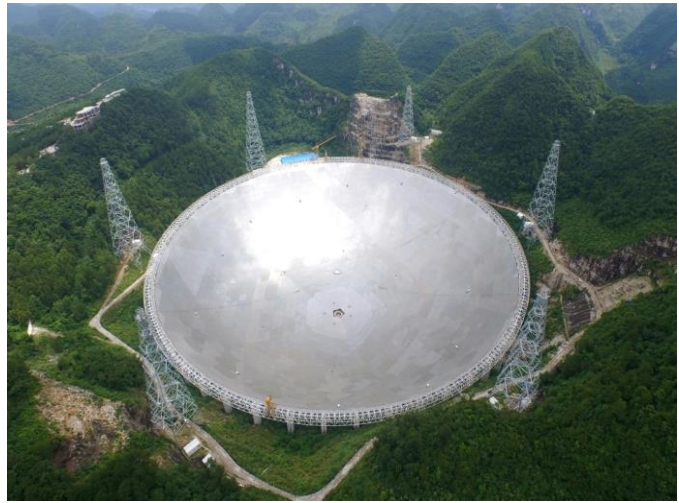

Fig. 1.1 Overview of the FAST telescope.

The illuminated aperture of the FAST is 300 m, and its overall performance and sensitivity are several times higher than those of the other existing radio telescopes. It is expected to remain the global leader in the field of radio telescopes for the next 20 years. On September 25, 2016, the FAST was completed with the main structure of the telescope installed, after which it entered the commissioning phase.

The FAST is a huge and complex mechanism with many influencing factors. It comprises 2,225 actuators and cable net, which forms a complex coupling active reflector system. A 30-ton feed cabin, which is driven by six cables, is about 140 m above the reflector. A failure of the feed support system or actuators may cause overall safety problems for the FAST. Therefore, the primary task of FAST commissioning is to improve its operational reliability and to ensure its effective observation time. Then, another effort has been made to improve the measurement and control the accuracy of the telescope for better efficiency, better pointing accuracy, and to reduce the system noise temperature.

Till date, several observation modes, such as tracking, drift scanning, and basketweave scanning, have been realized, which means that the functional commissioning tasks have been completed. Now, the sensitivity of the FAST has reached 2,000 m$^2$/K, the system noise temperature has been controlled below 20 K (19-beam 1.05-1.45 GHz receiver), and the pointing accuracy of the feed receivers has reached about 16″. The FAST has already begun to search for pulsars in batches.

Several researches based on the FAST telescope have been reported. Qian et al. [3] reported their observation and basic parameters of the first pulsar discovered by the FAST, PSR J1900−0134. Zhang et al. [4] used FAST parameters obtained from the commissioning data to estimate the sensitivity of the CRAFTS extragalactic HI survey and predict its survey capacity in the future. Yu et al. [5] observed the abnormal emission shift event of PSR B0919+06 using the FAST with the ultra-wideband receive system. They found the potential existence of a slow-drifting mode between two major abnormal events. A sequence of dimmed pulses was observed during one of those events at all frequency bands. Lu, Peng et al. [6] reported the analysis of three rotating radio transients (RRATs), namely J1538+2345, J1854+0306, and J1913+1330, observed using the FAST. The derived burst rates of the three RRATs are higher than previous results owing to the high sensitivity of the FAST. Lu et al. [7] showed both the mean and single pulses of PSR B2016+28, observed in detail using the FAST. Wang, Zhu, Guo, et al. [8] developed a pulsar candidate selection system based on deep neural networks and published the training candidate data that they collected from a drift scan pulsar search using the FAST wideband receiver.

This paper will introduce the implementation scheme, functional commissioning, key technical indices, and performance commissioning progress of the FAST.

## 2. Design Scheme and its Function Realization of FAST

### 2.1 Design scheme

The FAST is essentially a paraboloid antenna. With an innovative active reflector system and rigid–flexible hybrid lightweight feed support system, the FAST creates a new model for the construction of giant radio telescopes. The specific implementation plan is as follows:

**1) Active reflector system**

According to the geometric optical principle of the FAST, as shown in Fig. 2.1(a), the supporting structure of the reflector system should be capable of forming a parabolic surface from a spherical surface. As shown in Fig. 2.1(b), when the focal ratio is in the range of 0.46–0.47, the maximum deviation from the spherical surface to the paraboloid within a 300-m aperture is less than 1 m. Inspired by the support structure of the reflector of the Arecibo telescope, the FAST designers put forward the idea of using flexible cables to support the active reflector.

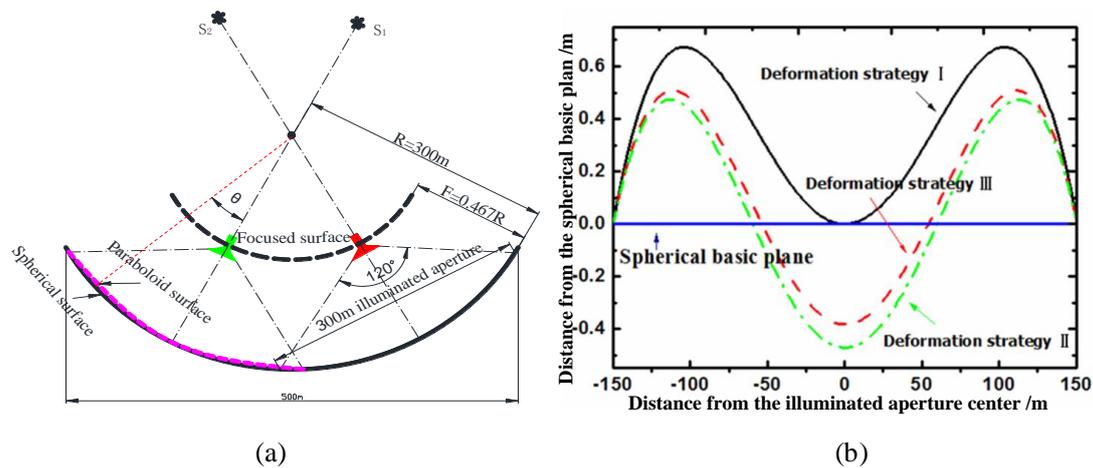

Fig. 2.1 Deformation distance analysis of spherical reflector to paraboloid: (a) geometric optical principle of the FAST and (b) deviation from the spherical surface to paraboloid within a 300-m aperture.[9]

The FAST cable net comprises 6,670 steel cables and ~2,225 cross nodes. The lengths of the cables range from 10.5 to 12.5 m, the total weight of the net is about 1,300 tons, and the main cables have 16 cross-sectional areas ranging from 280 to 1,319 $mm^2$.

As shown in Fig. 2.2, the outer edge of the cable net is suspended from a 500-m diameter steel ring beam. The cross nodes of the cable net are used as control points. Each of them is connected to an actuator by a down-tied cable. By controlling the actuator using feedback from the measurement and control system, the positions of the cross nodes can be adjusted to form a 300-m illuminated aperture. This illuminated aperture moves along the spherical surface according to the zenith angle (ZA) of the target objects (see S1 and S2 in Fig 2.1 (a)). The spherical reflector panel with a 500-m aperture is divided into 4,300 triangular units and 150 quadrilateral edge units for a total of 4,450 blocks. The border length of each triangular unit is about 11 m. The cable net acts as the supporting structure of the reflector units. The radius of curvature of the reflector units is about 315 m, and the working distance of the actuators is 1,200 mm, which satisfies the requirement of the reflector to deform from a spherical surface to a paraboloid [9–11].

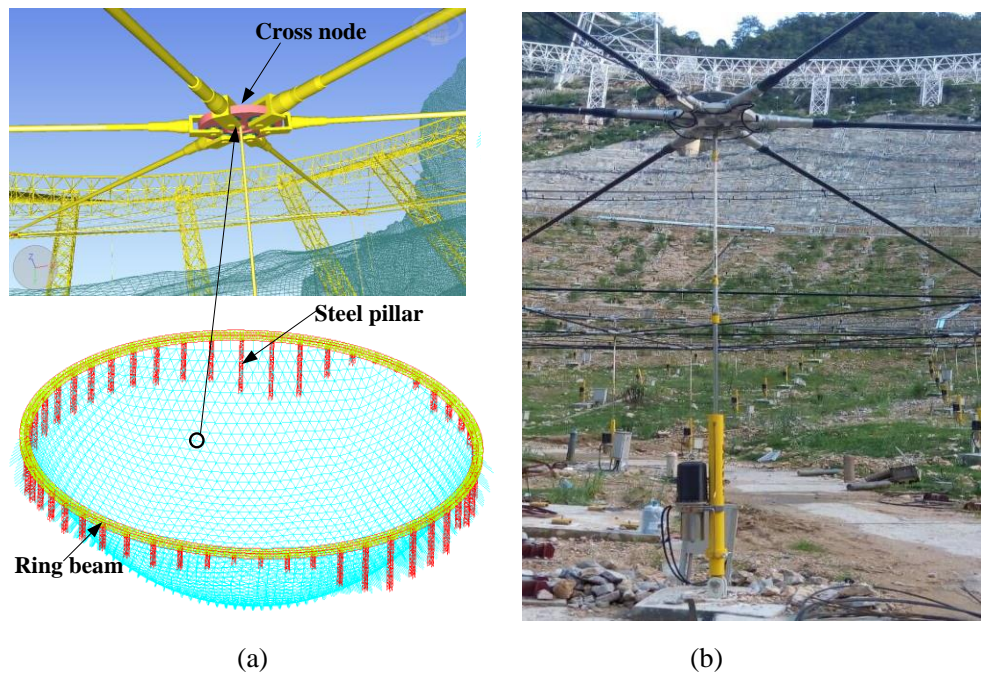

(a)                                    (b)

Fig. 2.2 Active reflector system: (a) ring beam and cable net model and (b) a cross node and its tie-down cable with actuator.

**2) Rigid–flexible hybrid lightweight feed support system**

With the reflector forming a paraboloid, the electromagnetic waves emitted from the celestial bodies can be converged onto the focal point. Maintaining the receiver at a correct position and orientation still remains challenging.

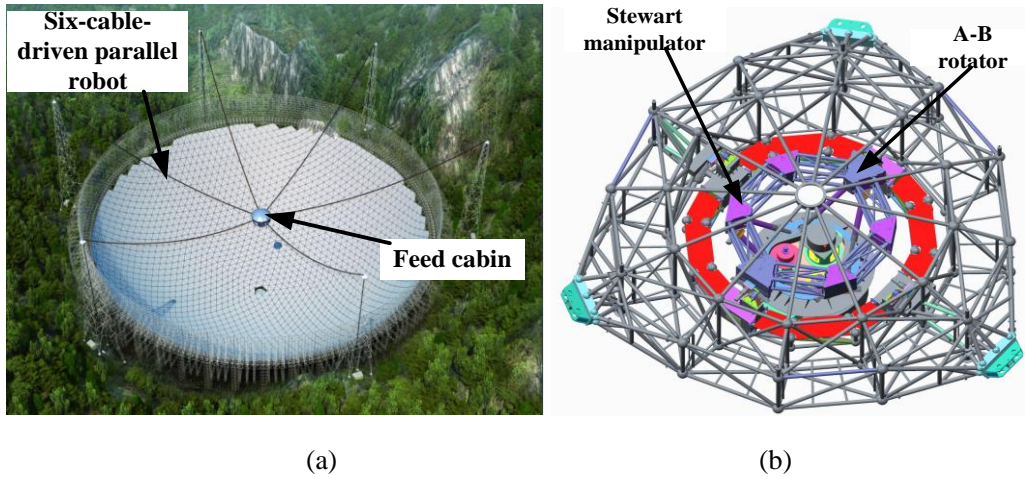

(a)                                                   (b)

Fig. 2.3 Design of the feed support system of the FAST: (a) main mechanisms of the feed support system and (b) main structure and mechanisms of the feed cabin.

As shown in Fig. 2.3(a), the FAST innovatively uses six cables, each of which is about 600 m long and weighs 7 tons. These cables parallel drive the feed cabin to move over a range of 206 m, at a height of 140 m above the reflector [12–16]. The cables and feed cabin are connected using three nodes. Each node uses a group of anchors, which greatly reduces cabin spin. The cables have a negligible shielding effect on the reflector and are of great benefit to further enhance the sensitivity of the telescope. At the same time, standing waves and other problems are also expected to be improved to achieve the broadband and full polarization function of the telescope.

However, it was found that the cable-driven parallel robot can control the feed receivers to their required position but cannot control them to their required orientation at the same time. Therefore, an A–B rotator, as shown in Fig. 2.3(b), which is a universal axis structure, is used to adjust the orientation of the feed receivers to the direction of the celestial source.

Through a simulation analysis of the long-span cable robot, it was found that the influence of wind disturbance was obvious. Therefore, a Stewart manipulator, as shown in Fig. 2.3(b), with six degrees of freedom (DOF) was adopted to reduce wind-induced vibration and further adjust the position and orientation of the feed receivers to operate with high accuracy.

The feed receivers are installed on the lower platform of the Stewart manipulator. Fig. 2.4 shows the 19-beam, 1.05-1.45 GHz receiver installed on the lower platform.

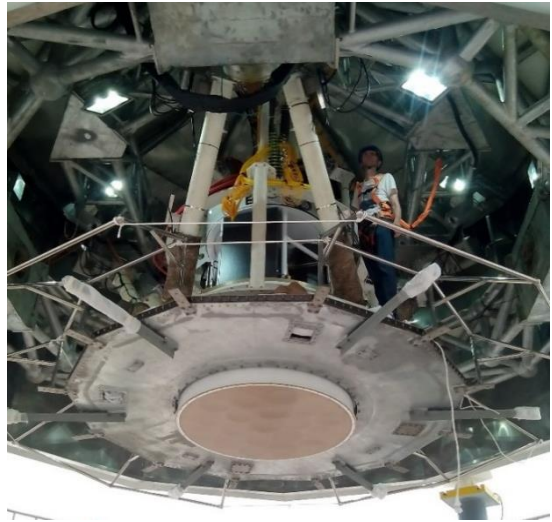

Fig. 2.4 The 19-beam 1.05-1.45 GHz receiver installed on the lower platform of the Stewart manipulator.

To realize closed-loop control, four targets each on the feed cabin and Stewart manipulator were used. Real-time measurement data are used to realize high-precision, closed-loop control of the feed support system.

## 2.2 Functional commissioning of the FAST

FAST telescope ensures safe operation of the active reflector system and feed support system owing to the real-time information system for the active reflector system and a hierarchical commissioning scheme for the feed support system. Finally, the safe operation of the two systems was realized, providing a guarantee for the performance commissioning of the telescope.

**1) Real-time information system of the active reflector system**

As the FAST starts operating, the shape of the cable net frequently switches between the 500-m spherical surface and the 300-m paraboloid. This causes cyclic loading and deformation of the ring beam and reciprocating movement of the reflector elements and restraining mechanism. Therefore, safety is always the basic problem of the active reflector system, including components such as the ring beam, cable net, reflector element, and accessory mechanism. Effective precautions are very important for preventing irreversible structural or mechanical damage caused by a malfunction or some unpredictable failure. Therefore, the real-time information system of the active reflector system is established to give timely information on cable stress and provide notification. It is an indispensable prerequisite for the routine operation of the telescope.

Fig. 2.5 shows the composition of the real-time information system of the active reflector system, which includes a data interface subsystem, data preprocessing subsystem, mechanical simulation model, and data post-processing subsystem.

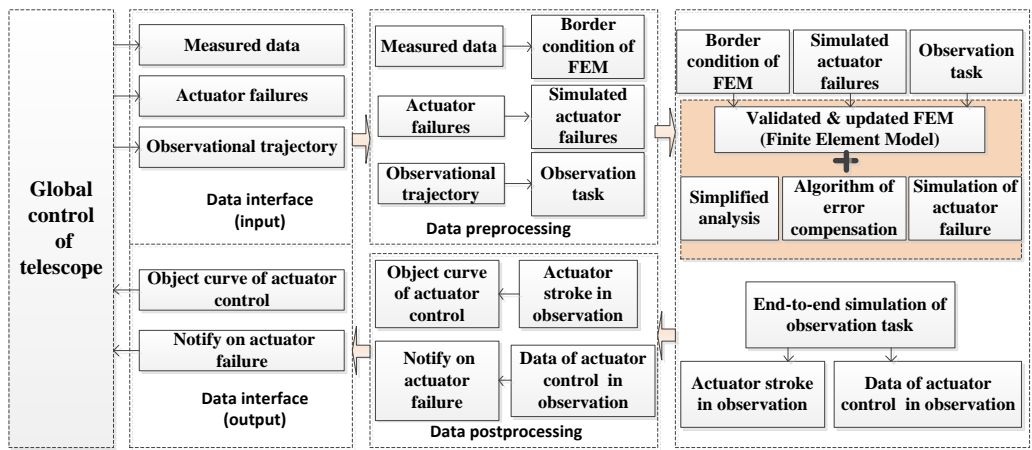

Fig. 2.5 Composition of the real-time information system of the active reflector system.

The data interface subsystem is the communication module between the overall system and the telescope control system (TSC). It receives the observational trajectory, sensor measurement information, and equipment malfunction information from the TCS and sends the appropriate notification information to the TCS.

The data preprocessing subsystem further translates the received information into data, working conditions, and constraints that can be identified by the simulation model to ensure the smooth progress of the simulation process.

The data post-processing subsystem sorts and analyzes the simulation results to identify possible dangerous structural stresses, translates the information into notification information that can be identified by the TSC system, and sends the notifications out in various ways to the system user interface to alert the operators.

The mechanical simulation model is the core of the real-time information system. It simulates all the key structures of the active reflector system, including the ring beam, lattice columns, cable net, reflector elements, and actuators. It is able to complete deformation computations of the cable net owing to its fast computation ability and output the resulting data once every 10-16s. In comparison with the deformation time (at least 300 s) of the cable net from the spherical shape to the paraboloid, the computational speed is quick enough for an assessment system. Deformation of the active reflector system is complicated and is influenced by many factors, such as temperature, actuator failure, side deviation of the cable nodes, and distortion of the tie-down cables. Some of them may even interact with one another. Fortunately, these factors and their interactions are easily simulated by finite element analysis; therefore, it is not necessary to analyze the contributions of individual factors. The remarkable feature of the simulation model is its real-time computation ability without establishing a database in advance and without considering the decoupling of miscellaneous aspects.

The real-time information system of the active reflector system worked well and achieved good results during the commissioning of the FAST. The system is very effective for the assessment of actuator failures and their impact. Additionally, the real-time information system can also judge whether there is danger caused by the paraboloid lying in the zone of large ZA, where the cable force is frequently found to be abnormal. These precautions remind telescope operators to remain vigilant to ensure the safe operation of the telescope.

**2) Functional commissioning of the feed support system**

The feed support system operates at more than 140 m above the reflector, so safety is the primary factor. Therefore, before performance commissioning of the feed support system, it was necessary to do some functional commissioning by stages, including the following:

a)  No-load commissioning of the six-cable-driven parallel robot

No-load commissioning was done by unloading the six cables from the reels and fixing the driving cables to the ground with splints to ensure no load on the overall mechanism. Then, the dynamic performance of the mechanisms of the robot was tested and debugged. When all the commissioning results meet the design indices, the commissioning ended.

b) Simulation commissioning of the six-cable-driven parallel robot

The simulation commissioning of the six-cable-driven parallel robot was to verify the operation process and control instructions of the robot. As shown in Fig. 2.6, based on the semi-physical simulation platform, all of the tests could be completed.

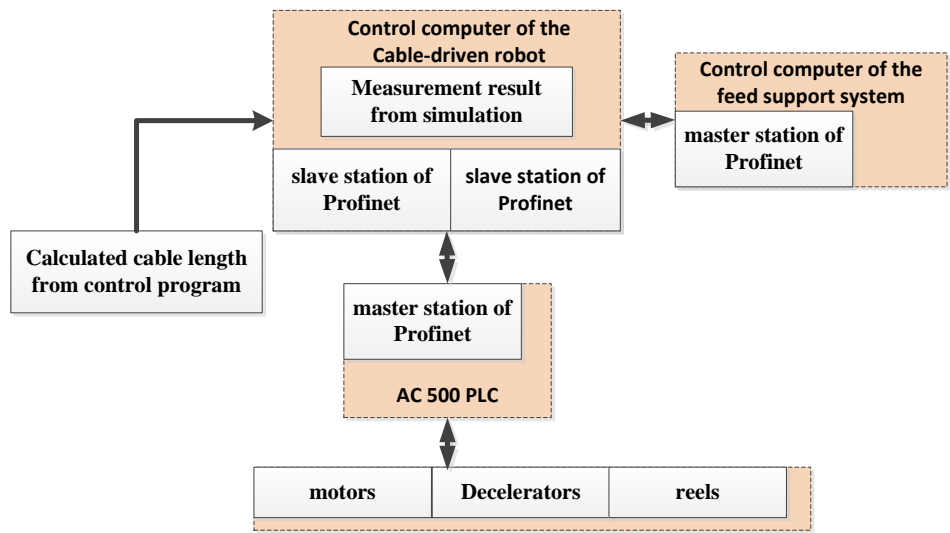

Fig. 2.6 Semi-physical simulation platform of the cable-driven parallel robot.

When the system started and the observation task was given, the control program calculated the driving cable length according to the current position and orientation of the feed cabin, and the results were sent to a programmable logic controller (PLC) through the process field net (PROFINET). Then, the results were given to the control computer by the user datagram protocol (UDP). The control computer sent all of the information to inversely solve the new position and orientation of the feed cabin. All of the simulation commissioning was done to ensure that the robot can perform routine, safe work.

c) Combined commissioning of the feed cabin and six-cable-driven simulation model

A feed cabin is a large mechanism with a diameter of 13 m and weight of 30 tons.

It is driven by six long-span cables. Before combined work with the robot, a semi-physical simulation model is adapted to debug the functions and performance of the feed cabin.

The semi-physical simulation model uses the main mechanism of the feed cabin and combines the robot's simulation model. The semi-physical simulation model has three features:

- The actual A-B rotator and Stewart manipulator as the physical part.
- Simulation software to simulate the coupling system of the cable and feed cabin under the condition of wind disturbance.
- Combined measurement means to simulate the actual measurement accuracy.

The concept for building the semi-physical simulation model is shown in Fig. 2.7.

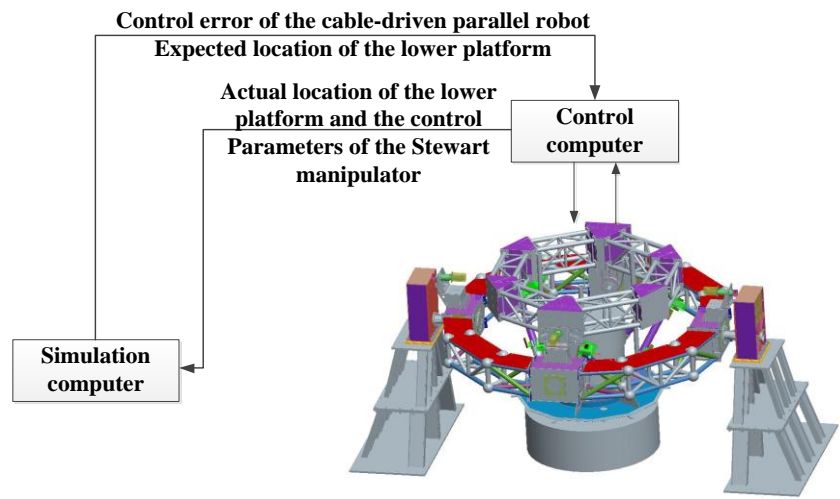

Fig. 2.7 Semi-physical simulation model.[16]

The semi-physical simulation model used the simulation system of the feed support system to generate wind disturbance parameters and simulate the control process and control error of the robot. The simulation system also gave the theoretical position and orientation of the lower platform of the Stewart manipulator, which were sent to control the feed cabin. These control parameters and the actual motor parameters were returned to the simulation computer, and the final accuracy of the feed support system were obtained by simulation. The semi-physical simulation model used a set of terminal measurement equipment, which simulated the period and accuracy of the real measurement system of the FAST. Therefore, the semi-physical simulation model verified the combined operation of the feed support system and provide a safe and reliable operating basis for the subsequent performance commissioning.

### 3) Functional commissioning results of FAST

Based on the functional commissioning mentioned above, the FAST realizes safe operation of the active reflector system and feed support system. On August 27, 2017, the FAST first continuously tracked a specific target source (3C286) twice; each time was about 10 min. This indicates that the most serious safety risks have been solved, which is a milestone for the telescope. By the end of 2017, the telescope had realized a variety of observation modes, such as tracking, drift scanning, and basketweave scanning, which completed its functional commissioning. Fig. 2.8 shows the observation results for November 17, 2017.

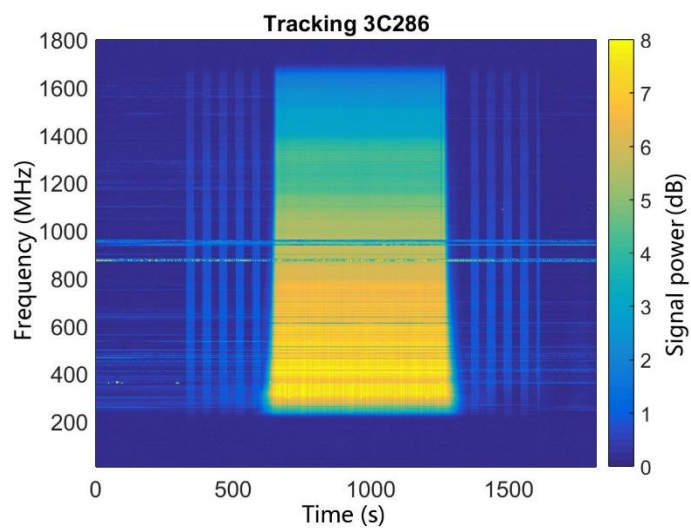

Fig. 2.8 Continuous trial observation of the radio source 3C286 on November 17, 2017.

## 3. Performance Commissioning of the FAST

FAST performance commissioning aims to test and optimize the parameters of the telescope so that it can meet the acceptance indices. The main commissioning work includes key technical index decomposition of the FAST and high-precision measurement and control commissioning of the active reflector system and feed support system.

### 3.1 Key technical indices and decomposition of the FAST

Table 3.1 lists the design and acceptance indices of the FAST. Currently, the purpose of the commissioning work is to achieve the acceptance indices, including pointing accuracy, sensitivity, and system noise temperature, among others.

Table 3.1 Planning technical and acceptance indices of the FAST

| Index Name | Planning Indices | Acceptance Indices (commissioning indices) |
|---|---|---|
| Diameter of the spherical reflector | 500 m | 500 m |
| Illumination aperture | 300 m | 300 m |
| Focal ratio | 0.46–0.47 | 0.46–0.47 |
| Sky coverage | ZA up to 40° tracking 4–6 h | ZA up to 40° tracking 4–6 h |
| Frequency band | 70 MHz–3 GHz | 70 MHz–3 GHz |
| Sensitivity | 2000 m$^2$/K | 1600 m$^2$/K |
| System noise temperature | 20 K | 25 K |
| Resolution (L-band) | 2.9' | 2.9' |
| Multi-beam (L-band) | 19 | 19 |
| Slewing | <10 min | <20 min |
| Pointing accuracy | 8" | 16" |

The FAST observation efficiency ($\eta$) is estimated to be

$$\eta = \frac{A_{eff}}{A_{geo}}, \tag{1}$$

where $A_{eff}$ is the effective illumination area and $A_{geo}$ is the geometric area of the telescope. For the FAST, $A_{geo}$ is the area of the 300-m diameter paraboloid.

As for the FAST efficiency, $\eta$ can be decomposed with several parameters that relate to a radio telescope:

$$\eta = \eta_{sf} \cdot \eta_{b1} \cdot \eta_s \cdot \eta_t \cdot \eta_{misc}, \tag{2}$$

where $\eta_{sf}$ represents the efficiency of the main reflector, relating to the accuracy of the reflector $\varepsilon$ and observational wavelength $\lambda$. It follows the *Ruze* equation, $\eta_{sf} = e^{-\left(\frac{4\pi\varepsilon}{\lambda}\right)}$, where $\eta_{b1}$ is the efficiency caused by the shielding of the feed cabin (an estimated value of one is adopted because of the small projected area of the feed cabin and its supporting structure relative to the reflector); $\eta_s$ is the spillover efficiency (a general value of 96% is adopted); $\eta_t$ is the illumination efficiency of the feed (about 76% when −13 dB Gaussian illumination is considered); $\eta_{misc}$ represents the efficiency of other aspects, such as the offset of the feed phase, matching loss of feed, among others). In this paper, the value of $\eta_{misc}$ is taken as 98% because we mainly consider the offset of the feed phase.

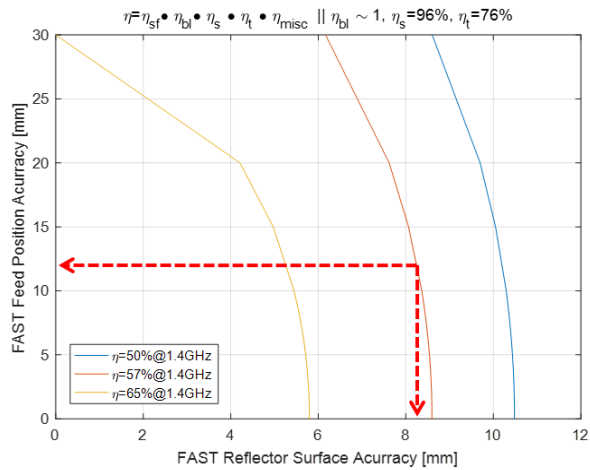

Fig. 3.1 Accuracy decompositions of the FAST.

By assuming a system noise temperature ($T_{sys}$) of 25 K, the FAST could satisfy 1600 m²/K of the sensitivity only if the efficiency of the telescope is at least 57%. Figure 3.1 shows the relations between the reflector and feed center accuracy for various telescope efficiencies. The red line indicates the reflector and feed center accuracies that should be met to realize an efficiency of 57%.

The telescope pointing accuracy was set as 16", which was converted to about root mean square deviation (RMS) 12 mm of the feed position measurement and control accuracy. According to Fig. 3.1, the corresponding surface precision of the reflector system is about RMS 8.2 mm. Further decompositions are as follows:

**1) Surface precision of the reflector system**

Based on the characteristics and measurement method of the active reflector system, an open-loop control system is adapted for the active reflector system's deformation. The surface precision of the reflector system is affected mainly by some known errors, such as panel design errors (RMS 2.2 mm), panel manufacturing errors (RMS 2.5 mm), panel temperature effect errors (RMS 1 mm), panel wind load errors (RMS 1 mm), and installation errors (RMS 1 mm), among others.

The known error caused by the panels is estimated to be about RMS 3.75 mm, which was already determined in the design of the construction phase. These errors would directly accumulate into the surface error of the reflector system.

According to the long-term measurement data and simulation results, the surface error caused by temperature ranges from RMS 0.91 to RMS 4.13 mm with the different temperature (±20°C) and different ZA (0°–26.4°). In most cases, this error is no more than RMS 2 mm. If the database model of the control system is well established in the future, the surface error caused by temperature is expected to be reduced to less than RMS 1 mm. However, it is reasonable to directly accumulate the average of this to the overall system; thus, it is recorded as RMS 2 mm.

The measurement error of the active reflector is influenced mainly by spherical aberration, measurement shafting deviation, atmospheric refraction, and the accuracy of the measurement reference net. By establishing an accurate correction model for spherical aberration, it is predicted that its influence can be reduced to less than RMS 1 mm. The horizontal and vertical axes of the total station are not orthogonal; therefore, measurement shafting deviation is the deviation of the horizontal and vertical axes of the total station. The shafting deviation after calibration can be limited to 1", and the maximum distance of the side length of the reference net is ~200 m, which corresponds to RMS 0.8 mm. By precise difference, the error caused by atmospheric refraction is expected to reduce to 1 mm and that of the measurement reference net can also be reduced to 1 mm. After accumulative analysis and with a certain allowance, the measurement accuracy index of the active reflector system is set as RMS 2 mm.

According to the FAST condition, we have developed a new open-loop control method for reflector shape control, wherein real-time shape measurement is not otherwise necessary. The details of the open-loop control method are illustrated in Section 3.3. Here we assume that the control precision index of the cable net can be kept within RMS 3.5 mm.

Based on the abovementioned error analysis and decomposition, the final surface precision of the reflector system is expected to achieve RMS 6 mm. To achieve the accuracy index, two prerequisites are important: 1) a high-precision, open-loop control system for the active reflector system and 2) a high-precision measurement system for the active reflector system. Therefore, the key objective of the active reflector system commissioning is the realization of these two prerequisites.

**2) Accuracy of the feed support system**

The accuracy index of the feed pointing of the FAST is 16", and the corresponding measurement and control accuracy of the feed support system is RMS 12 mm. According to the general requirements of dynamic control theory, the measurement accuracy must reach about one-third of the overall measurement and control accuracy of the system, i.e., RMS 4 mm.

After analyzing the measurement accuracy of the feed support system, it was found that it was difficult to greatly reduce the angle measurement error of the total station for a large range of motion and long measurement distance of the feed. Therefore, a method of distance intersection was adopted, which only used the distance information of the total station. By analyzing and calculating the distance information of the total station, the position and orientation of the feed cabin and Stewart manipulator were obtained.

The measurement errors of the feed support system come mainly from the error of the measurement reference net, target error of the total station, the random error of the total station, and time delay. The precision of the measurement reference net is better than 1 mm. Through analysis and estimation, the target error of the total station is about RMS 1 mm. According to the performance parameters of the total station, the random error in the whole workspace of the feed support system can be kept within RMS 3 mm. The total delay time of the total station system is about 100 ms, which would lead to an error of ~RMS 1.2 mm. After a comprehensive analysis, the measurement error of the feed support system will be within RMS 4 mm.

Based on the error decomposition, the measurement and control indices of the active reflector system and feed support system are given. After that, the performance commissioning aiming at the technical indices is carried out for the telescope.

### 3.2 Commissioning of the measurement system

**1) Measurement reference net**

The measurement system is an important part of the FAST telescope, and its performance directly affects the observation performance of the telescope [17, 18]. The accuracy of measurement system is affected mainly by the measurement station distribution and accuracy, environmental factors, and measuring equipment factors, among others. The influence of the environment and measuring equipment can also be analyzed and estimated. For the impact of the measurement stations, we set up a measurement reference net, which achieves 1 mm accuracy over a range of 500 m. The specific implementation methods are as follows:

As shown in Fig. 3.2, 23 measurement stations are established as the FAST reference measurement net, which is the basis of the measurement system and supplies precision measurement stations for the feed support and active reflector systems. The core problem of precise measurement of the FAST is how to measure the exact position of these reference stations and realize millimeter-level precision. For the high-precision calibration and periodic stability monitoring of the FAST measurement reference net, an auto-monitoring system was developed.

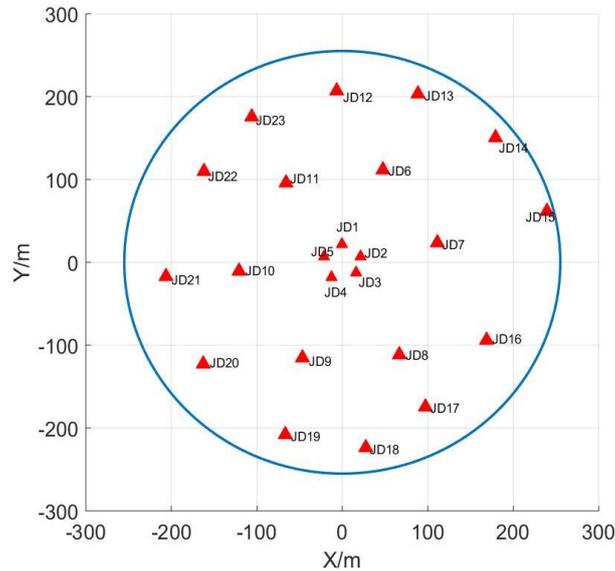

Fig. 3.2 Position distribution of the measurement foundations.

Based on the precision decomposition, the coordinate precision of the measurement reference net should be better than 1 mm. Considering the requirement of the high-precision calibration and periodic stability monitoring of the FAST measurement reference net, the single-measurement duration of the measurement reference net should be better than 2 h to ensure the consistency of atmospheric properties, such as temperature, humidity, and air pressure, among others.

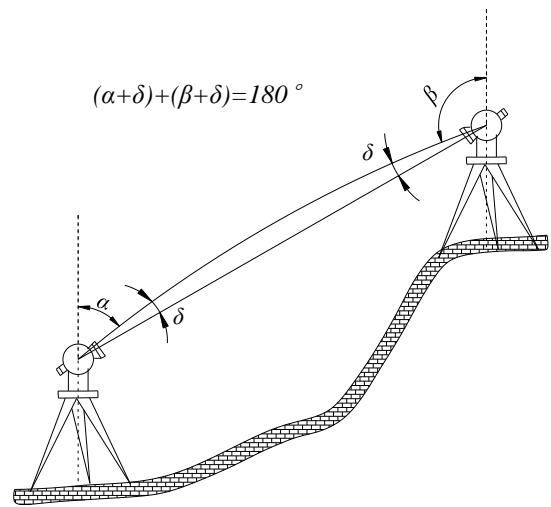

Fig 3.3 Mutual aiming technology of the total station.

To achieve the high-precision measurement requirements, the telescope uses mutual aiming technology of the total station. As shown in Fig 3.3, high-precision bi-prism devices are installed on two total stations on the same observation path. Both the stations mutually aim and measure each other's prisms, so that four sets of distance and angle measurement information are obtained and their average solution is calculated as the measurement information of the observation path. As illustrated by Fig. 3.3, the sum of the altitude angles measured by the adjacent total stations can be expressed as:

$$(\alpha + \delta) + (\beta + \delta) = 180°. \tag{3}$$

Thus,

$$\delta = \frac{180° - \alpha - \beta}{2}. \tag{4}$$

On one hand, this technology reduces most of the angle measurement errors caused by atmospheric influences. On the other hand, it offsets the error of the total station's coordinate system and the installation error of the bi-prism devices.

Observation planning paths (115) are obtained, as shown in Fig. 3.4, and the mutual aiming technology is used for all the total stations on the planning paths.

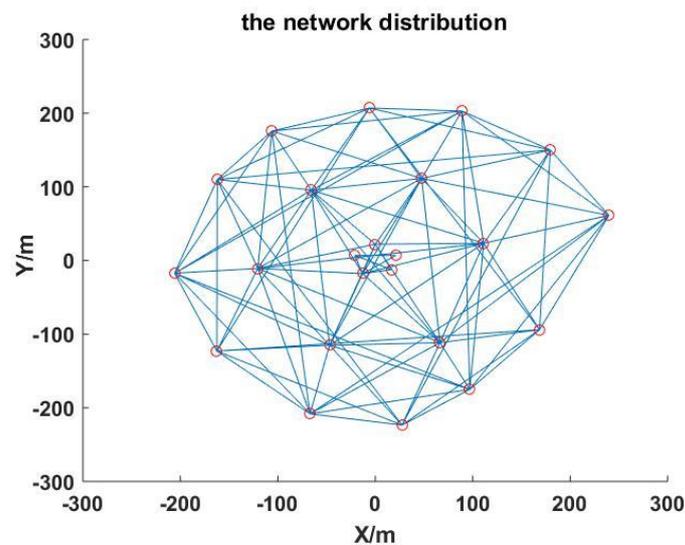

Fig. 3.4 Observation control net (measurement foundation Nos. 1 to 23).

After analyzing and dealing with the significant error, to improve the measurement accuracy of the measurement reference net, the FAST performs the elevation and plane net adjustments. The adjustment processing steps are as follows [19].

The observation vectors consist of $n$ independent observations, and the corresponding weight matrix $\boldsymbol{P}$ is a diagonal matrix. The modifying vector is $\boldsymbol{V}_{n\times 1} = [v_1, v_2, \cdots, v_n]^{\mathrm{T}}$, and the adjustment vector is $\boldsymbol{L}_{n\times 1}$. When there are $r$ redundant observations, the adjustment should satisfy $r$ adjustment equations:

$$\begin{cases} a_{11}\hat{L}_1 + a_{12}\hat{L}_2 + \cdots + a_{1n}\hat{L}_n + C_1 = 0 \\ a_{21}\hat{L}_1 + a_{22}\hat{L}_2 + \cdots + a_{2n}\hat{L}_n + C_2 = 0 \\ \cdots \\ a_{r1}\hat{L}_1 + a_{r2}\hat{L}_2 + \cdots + a_{rn}\hat{L}_n + C_r = 0 \end{cases}, \tag{5}$$

where $\hat{L}_i = L_i + v_i$, $(i = 1, 2, \cdots, n)$.

Let $[w_1, w_2, \cdots, w_r]$ be the closing difference between the conditional equations, then $r$ inconsistent equations are

$$\begin{cases} w_1 = a_{11}L_1 + a_{12}L_2 + \cdots + a_{1n}L_n + C_1 \\ w_2 = a_{21}L_1 + a_{22}L_2 + \cdots + a_{2n}L_n + C_2 \\ \cdots \\ w_r = a_{r1}L_1 + a_{r2}L_2 + \cdots + a_{rn}L_n + C_r \end{cases}. \tag{6}$$

We can get $r$ modified condition equations

$$\begin{cases} a_{11}v_1 + a_{12}v_2 + \cdots + a_{1n}v_n + w_1 = 0 \\ a_{21}v_1 + a_{22}v_2 + \cdots + a_{2n}v_n + w_2 = 0 \\ \cdots \\ a_{r1}v_1 + a_{r2}v_2 + \cdots + a_{rn}v_n + w_r = 0 \end{cases}. \tag{7}$$

Therefore, Eqs. (5), (6), and (7) can be expressed as

$$\boldsymbol{A}\hat{\boldsymbol{L}} + \boldsymbol{C} = 0, \tag{8}$$

$$\boldsymbol{W} = \boldsymbol{AL} + \boldsymbol{C}, \text{ and} \tag{9}$$

$$AV + W = 0, \tag{10}$$

where

$$A = \begin{bmatrix} a_{11} & a_{12} & \cdots & a_{1n} \\ a_{21} & a_{22} & \cdots & a_{2n} \\ & & \vdots & \\ a_{r1} & a_{r2} & \cdots & a_{rn} \end{bmatrix}, \; L = \begin{bmatrix} L_1 \\ L_2 \\ \vdots \\ L_n \end{bmatrix}, \; \hat{L} = \begin{bmatrix} \hat{L}_1 \\ \hat{L}_2 \\ \vdots \\ \hat{L}_n \end{bmatrix}, \; C = \begin{bmatrix} C_1 \\ C_2 \\ \vdots \\ C_r \end{bmatrix}, \; \text{and} \; W = \begin{bmatrix} w_1 \\ w_2 \\ \vdots \\ w_r \end{bmatrix}$$

Since the total number of observations ($n$) is greater than the number of redundant observations ($r$), i.e., $n > r$, there are infinite groups of solutions for Eq. (10). According to the principle of adjustment, taking the values $V^{\mathrm{T}}PV$ that are equal to the minimum set of values of $V$, we set the connection number vector $K^{\mathrm{T}} = [k_1, k_2, \cdots, k_r]$ and the function $\Phi$ as

$$\Phi = V^{\mathrm{T}}PV - 2K^{\mathrm{T}}(AV + W). \tag{11}$$

By seeking $\Phi$, the first partial derivative by $V$ and making it 0, we can obtain

$$\frac{\partial \Phi}{\partial V} = 2V^{\mathrm{T}}P - 2K^{\mathrm{T}}A = 0 \; \text{and} \tag{12}$$

$$V^{\mathrm{T}}P = K^{\mathrm{T}}A. \tag{13}$$

The correction formula is obtained from Eq. (13) as follows:

$$V_{n \times n} = P^{-1}_{n \times n} A^{\mathrm{T}}_{n \times r} K_{r \times 1}. \tag{14}$$

The elevation net adjustment adopts the free net adjustment with additional zenith distance parameters. As shown in Fig. 3.5, the standard deviation of the elevation residual error is RMS 0.2 mm.

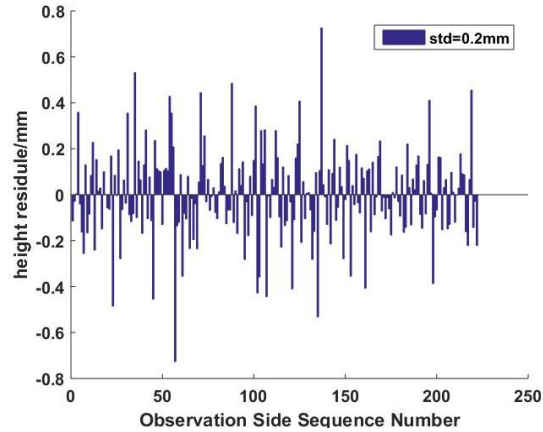

Fig. 3.5 Elevation residual error.

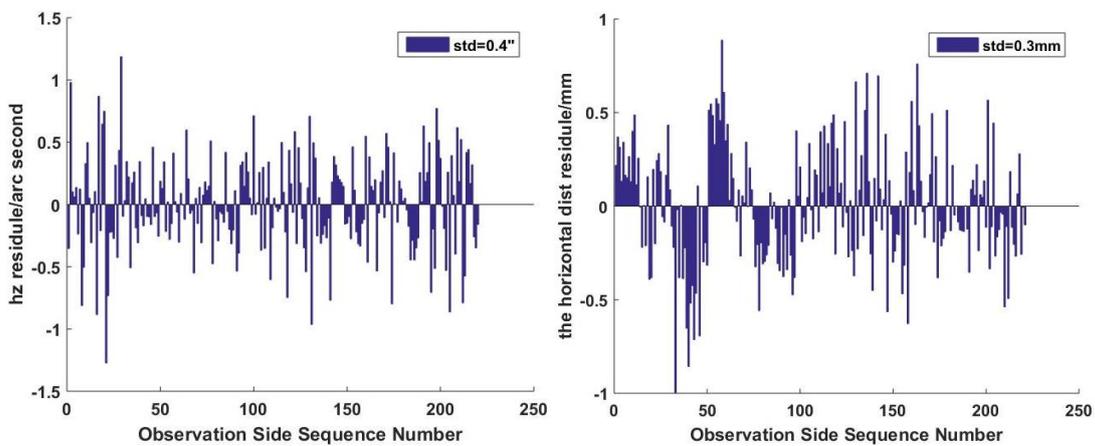

Fig. 3.6 Horizontal angle residual error.　　Fig. 3.7 Horizontal distance residual error.

The plane net adjustment selects the No. 4 measurement foundation as the known reference measurement and the azimuth of the Nos. 4 to 9 measurement foundations as the known azimuth angle. The free net adjustment with additional prism constants or instrument constants is applied to the plane net adjustment. As shown in Figs. 3.6 and 3.7, the standard deviations of the horizontal angle residual and horizontal distance residuals are 0.4" and 0.3 mm, respectively, and the plane coordinate error of the 23 measurement foundations is RMS 0.2 mm.

The observation results show that the elevation coordinates measurement accuracy is better than RMS 0.1 mm, the plane coordinate measurement accuracy is better than RMS 0.2 mm, and the measurement time is less than 30 min once, which is better than the requirement of 1-mm coordinate measurement accuracy and 2-hour measurement efficiency.

**2) Measurement method of the active reflector system and feed support**

**system**

After the accuracy decomposition of the FAST, the measurement accuracy index of the active reflector system is set as RMS 2 mm, and the measurement error of the feed support system should be within RMS 4 mm.

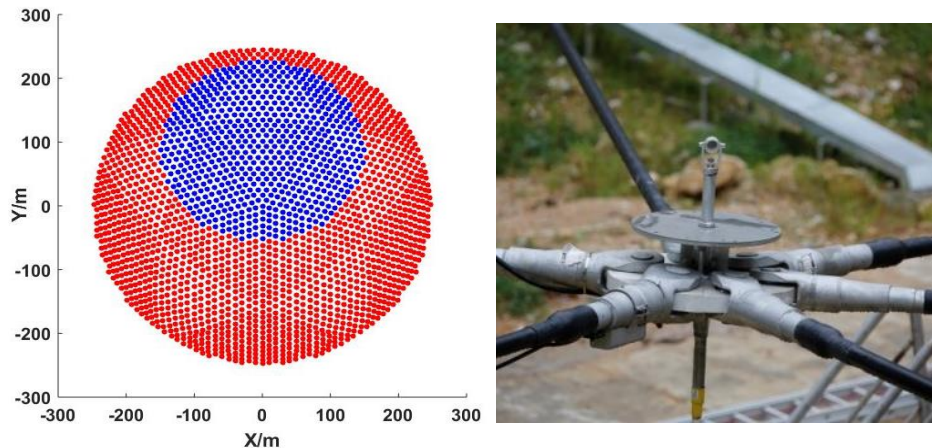

Fig. 3.8 Distribution of the total station target.

The measurement system of the active reflector system uses ten total stations installed on five measurement foundations on the inner circle of the FAST reference measurement net. As shown in Fig. 3.8, the total stations measure the 700 targets on the paraboloid and 2,225 targets on the entire cable net, used for static calibration and to provide measurement data for the open-loop control system of the active reflector system.

Currently, the measurement system of the active reflector system is used mainly for static surface calibration. Through correction of the total stations' coordinate system error, the correction for the effect of the Earth's curvature and refraction, and the unification of coordinate systems, the consistency error of the total stations is controlled below RMS 1.5 mm. The paraboloid error is about RMS 1.4 mm, which satisfies the accuracy requirements of the paraboloid reflector.

The measurement system of the feed support system currently uses four total stations installed on the measurement foundations on the outer and middle circles. Four laser measuring targets are installed on the feed cabin and on the lower platform of the Stewart manipulator. Through the real-time measurement data of the FAST measuring net, the precise positions and orientations of the feed cabin and Stewart manipulator are solved, which provides high-precision measurement data for controlling the feed support system.

### 3.3 High-precision control realization of the active reflector system and feed support system

As shown in Fig. 3.9, an open-loop control system for the active reflector system was developed. The open-loop control system first needed to build a calibration database. Initially, a very precise reference sphere is formed by repeated measurement and feedback control, and then the position information of all the actuators at that moment is recorded. All of the information is used to modify the integrated mechanical model of the reflector element, cable net, and ring beam. Meanwhile, factors, such as temperature, actuator failure, side deviation distance, and tie-down cable deformation, are analyzed in the model. The accuracy of the mechanical model depends on two factors: 1) having a reasonable method of modeling and simplification; 2) measurement information from the high-precision sensor used to update the simulation model. Then, a large-scale mechanical simulation is conducted to obtain the database of all actuator location information related to various paraboloid and ambient temperatures. Finally, using the interpolation method, the target stroke and change curve of the actuator are obtained, which can be used for continuous tensioning of the cable net. The incremental positions of the actuators for continuous deformation control of the active reflector can then be calculated by mathematical interpolation and sent to the PLC for execution.

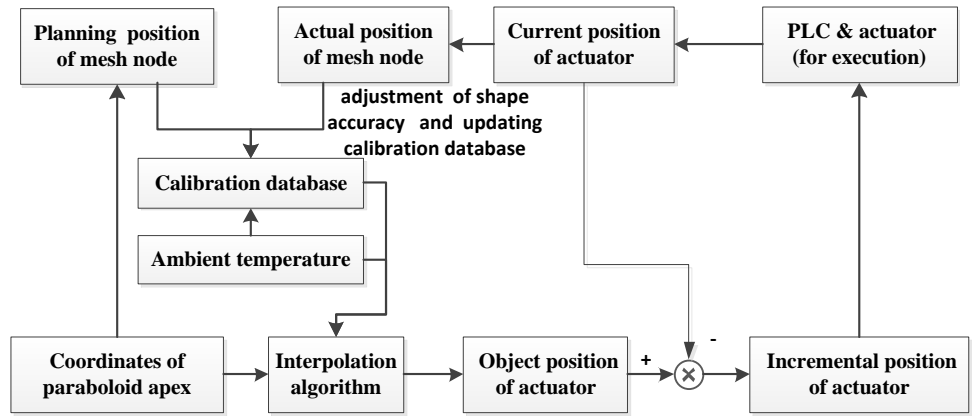

Fig. 3.9 Open-loop control system of the active reflector system.

The open-loop control system of active reflector system of the FAST comprises hardware and software parts, including four functional modules, which are a three-dimensional calibration database of cable net deformation (including the calibration value of the reference spherical surface), a query and retrieval module of the calibration database, an interpolation algorithm module, and a calibration module.

The open-loop control system of the active reflector system has a series of advantages. First, it eliminates the dependence on the real-time measurement to meet the requirement of all-weather operation, and it does not produce radio frequency interference(RFI). Also, in the calibration process, the safety problem of a cable net with paraboloid deformation at large ZA can be solved at the same time, so that every calibrated paraboloid is safe for the entire structure. In addition, using a high-performance computer to establish an efficient query and retrieval algorithm of the database and interpolation algorithm, the calculation time can fully meet the 0.5–1 s control cycle. Finally, by constructing the high-precision paraboloid calibration database, the interpolated paraboloid can also meet the control accuracy requirements when a certain number and density are reached.

Fig. 3.10 shows the surface accuracy of the active reflector system. After the establishment of the interpolation algorithm and calibration, the surface error of the basic spherical surface is less than RMS 1.7 mm. We changed the spherical surface to a paraboloid and measured the surface error of the reflector. Fig. 3.11 shows four

**25** / **49**

different paraboloids and their surface errors. From Fig. 3.11, it can be seen that the surface error of the paraboloid is less than RMS 2.6 mm, which satisfies the RMS 3.5 mm requirement.

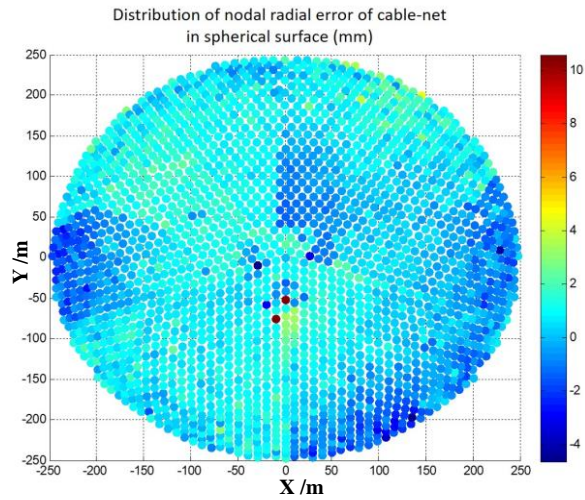

Fig. 3.10 Surface error of the basic spherical surface.

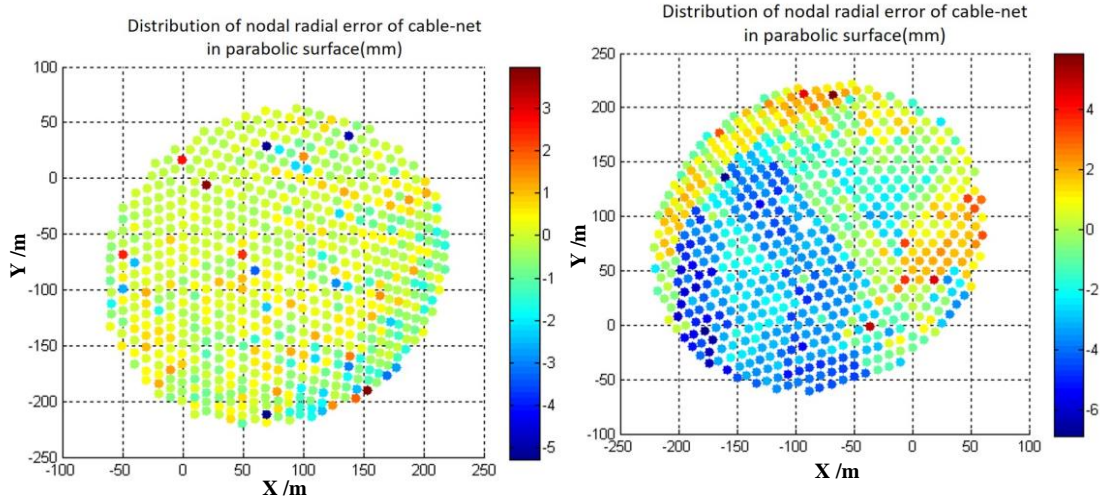

(a) Surface error is RMS 0.8 mm     (b) Surface error of is RMS 2.6 mm

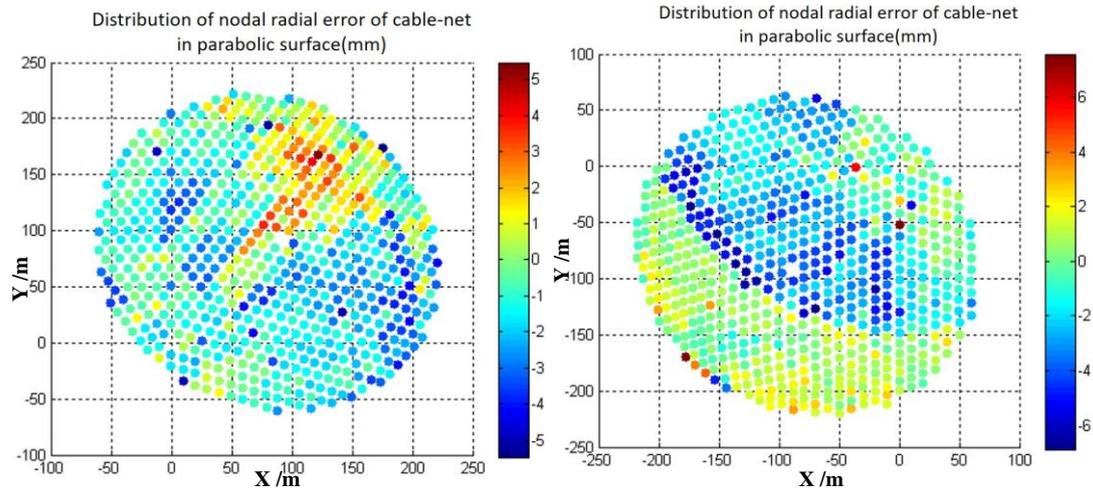

(c) Surface error is RMS 1.9 mm    (d) Surface error is RMS 2.3 mm

Fig. 3.11 Surface errors of the four paraboloids.

The feed support control system needs to control the feed and receivers installed in the feed cabin on a focal point within RMS 10-mm positioning accuracy in real time. To meet the accuracy requirement, a combined method of open- and closed-loop control is adopted. The pure proportional element of a closed-loop control will lead to a steady-state control error when a ramp input is used. An integrating element can compensate for the error but it can also bring a noticeable vibration to the flexible cable system. Besides, adjustment of an integral element is very slow and cannot be used effectively. Therefore, on the basis of the closed-loop control, open-loop control is added. According to the trajectory planning, the cable speed is sent directly to the motor for execution, which is also called the speed feed forward.

The feed support control system is divided into three active mechanisms: the six-cable-driven parallel robot, A-B rotator, and Stewart manipulator. According to the planning position and orientation of the feed and receivers, all the control parameters can be calculated. Then, each mechanism completes different control tasks. The control strategy is shown in Fig. 3.12.

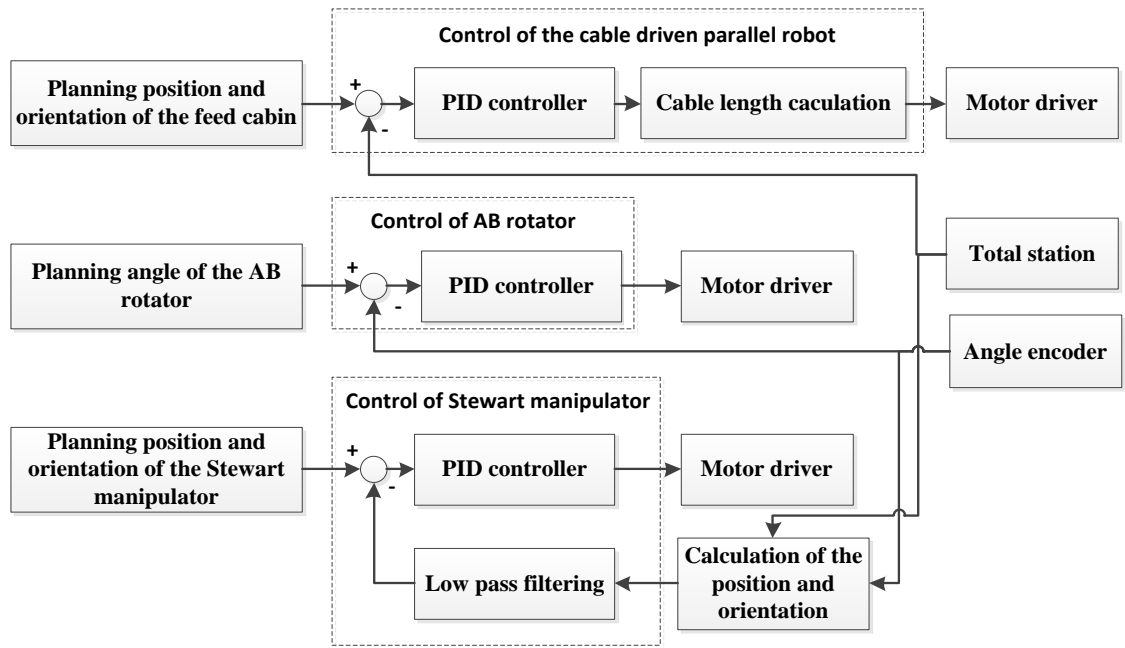

Fig. 3.12 Control strategy of the feed support system.

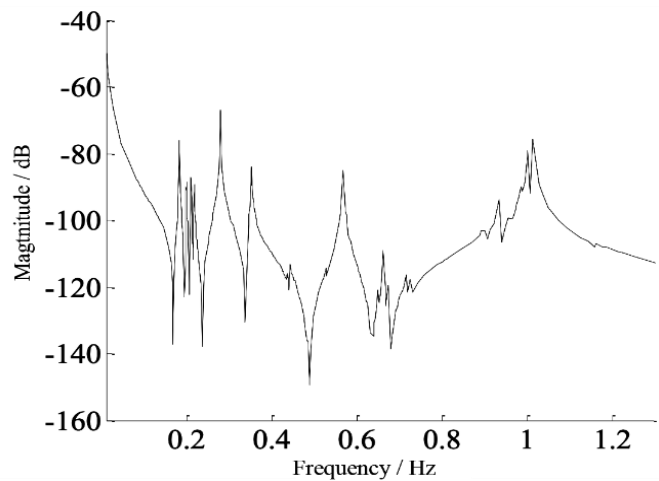

Fig. 3.13 Natural frequency of the cable-driven parallel robot.

Fig. 3.13 shows very low natural frequency and damping of the cable suspension finite element model. As a result of parameter identification of the real structure, the damping ratio is 0.2% and the first-order natural frequency is about 0.18 Hz. The cable–cabin system is easily excited by wind loads, which are difficult to dissipate quickly. Therefore, although the cable-driven parallel robot uses closed-loop control, it can only compensate for the steady-state error of the position of the feed cabin. Meanwhile, a cable tension uniform control strategy is adopted to ensure the optimal control response of the feed cabin in its workspace.

The A-B rotator comprises two high-stiffness ring frames, which adjust the pointing of the feeds by rotating the Stewart manipulator. This rotator adopts an open-loop control method using an absolute angle encoder to realize high control accuracy.

A Stewart manipulator, as a six-DOF parallel mechanism, has high stiffness, high kinematic accuracy, and good dynamic response. Using the measurement results of the total station, the position and orientation of the lower platform of the manipulator are obtained in real time, and then the accurate positioning of the feed receiver can be realized by the closed-loop control. From May to November, 2017, the software function, tracking accuracy, and control system parameters of the feed support system were debugged. Finally, the feed support system met the requirements in more than 30 randomly selected observation trajectories.

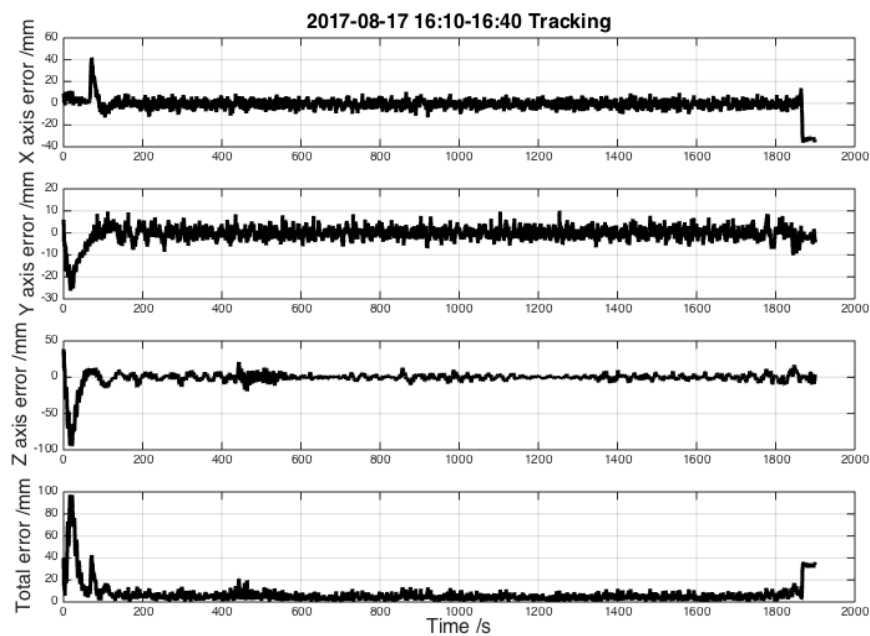

Fig. 3.14 Experiment results in observation of the feed support system.

Fig. 3.14 shows that the control error of one tracking trajectory is RMS 5.4 mm and the orientation error is RMS 0.3°. Considering that the measurement error is about RMS 4 mm, the overall measurement and control error of the feed phase center can be limited to RMS 7 mm to meet the requirement.

## 4. Performance Testing of the FAST

### 4.1 Performance test of the receivers and their backend

The receivers and backend are important components of a radio telescope. They are constantly being updated to serve new scientific goals and may largely improve the telescope's performance and functionality.

**1) The use of a low-frequency, ultra-wideband receiver and 19-beam 1.05-1.45 GHz receiver**

Since the commissioning of the FAST, two sets of the receivers have mainly been used and comprehensively tested: a low-frequency, ultra-wideband receiver and a 19-beam 1.05-1.45 GHz receiver.

The low-frequency, ultra-wideband receiver was jointly developed by the National Astronomical Observatory of the Chinese Academy of Sciences and the California Institute of Technology (https://ieeexplore.ieee.org/stamp/stamp.jsp?arnumber=7734751.) It covers the frequency range from 270 to 1,620 MHz and was mainly used for early scientific observations and commissioning of the FAST. This receiver has been used for the pointing test, beam test, and tracking scan of the FAST telescope.

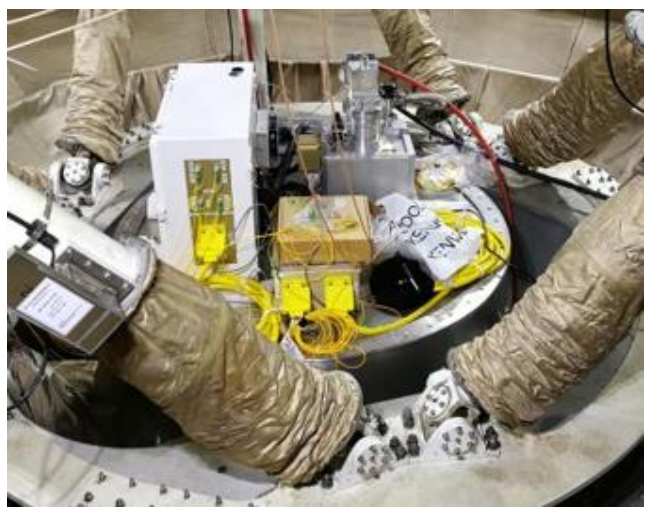

Fig. 4.1 Low-frequency, wideband receiver (Dewar and warm electronics) on the telescope feed cabin.

As shown in Fig. 4.1, the wideband receiver comprises three main parts: a Quad Ridge Flared Horn (QRFH), cryogenic receiver, and normal temperature receiver. The QRFH feed adopts a square structure, which is shown in Fig. 4.2. Owing to its low-frequency coverage (as low as 270 MHz), the size of the QRFH is up to 1.45 m$^2$ × 1.2 m; therefore, it cannot be cooled by a Dewar and it operates at room temperature. In its passband, its two-port reflection losses are greater than 10 dB except for the frequency band below 300 MHz. The cryogenic receiver includes mainly microwave components, such as the cryogenic Dewar, low-noise cryogenic amplifier, cryogenic filter [20], and cryogenic noise diodes, among others. The first and second cold heads of the cryogenic Dewar can cool to 10 and 50 K, respectively. The normal temperature receiver includes components, such as a normal temperature radio frequency (RF) amplifier, filter, attenuator, control circuit, and optical transmitter.

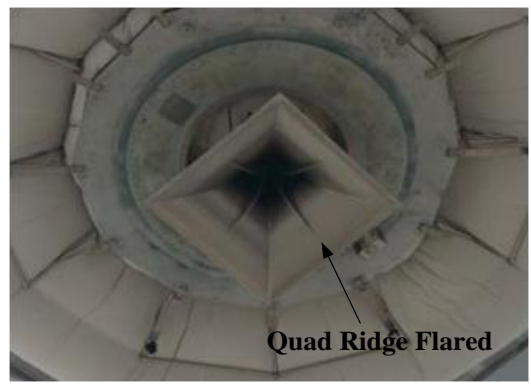

Fig. 4.2 QRFH of the wideband receiver on the FAST.

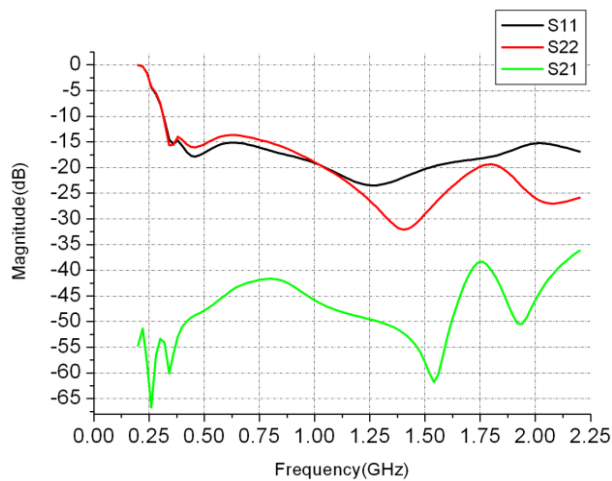

Fig. 4.3 Return loss of the QRFH of the FAST wideband receiver.

The low-frequency, ultra-wideband receiver was installed in the feed cabin in September 2016; its test results are shown in Fig. 4.3. In May 2017, this receiver realized cryogenic observation, and the first and second cold heads of the Dewar were cooled to 49 and 9 K, respectively. Referring to the input port of the cryogenic Dewar, the equivalent noise temperature is lower than 17 K, of which the low-noise amplifier placed on the Dewar's secondary cooling plate contributes 6 K. The remaining noise temperature is introduced by the coaxial line on the fore-end of the cryogenic amplifier.

The normal temperature receiver realizes RF signal amplification, filtering, receiver gain adjustment, and other functions. It contains H and V polarization channels, each of which is divided into broadband and narrowband bandpass signals, covering the frequency ranges of 270–1,620 MHz and 1,300—1620 MHz, respectively. The front end of the normal temperature receiver also includes a linear–circular polarization converter, which realizes the linear and circular polarization reception of the receiver. At the backend of the normal temperature receiver, four-channel RF signal optical transmitters are installed to transmit the astronomical signals from the front end to the observation room by optical fiber. Since the low-frequency, ultra-wideband receiver began operation, more than 60 high-quality pulsar candidates have been observed, 54 of which have been confirmed.

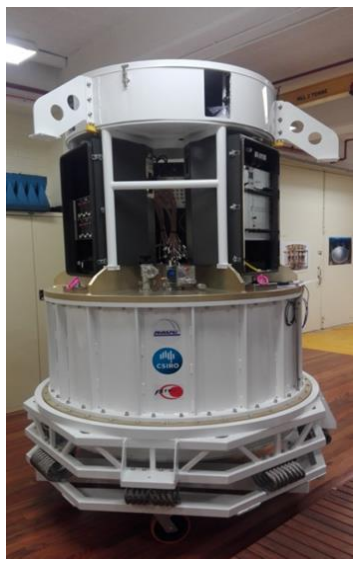

Fig. 4.4 Assembled FAST-CSIRO 19-beam 1.05–1.45 GHz receiver in the laboratory.

In July 2018, the 19-beam 1.05-1.45 GHz receiver, as shown in Fig. 4.4, was installed on the FAST telescope and began astronomical observations and commissioning. The receiver was developed jointly by the Australian CSRIO and the National Astronomical Observatory, which comprises 19-beam 1.05-1.45 GHz and covers a frequency range from 1.05 to 1.45 GHz (http://www.atnf.csiro.au/technology/receivers/FAST_Multibeam.html). In Fig. 4.5, the blue curve shows the output bandpass of the warm electronics and the purple curve shows that of the optical receiver. As shown in Fig. 4.5, the 19-beam 1.05-1.45 GHz optical transmission system provides a gain of ~15 dB, which greatly enhances the FAST telescope's observation ability. This receiver uses three cold heads to cool the inner microwave device of the Dewar. The temperatures of the first and second cold plates are 17 and 58 K, respectively, and the main cryogenic components include an orthogonal polarizer and cryogenic amplifier.

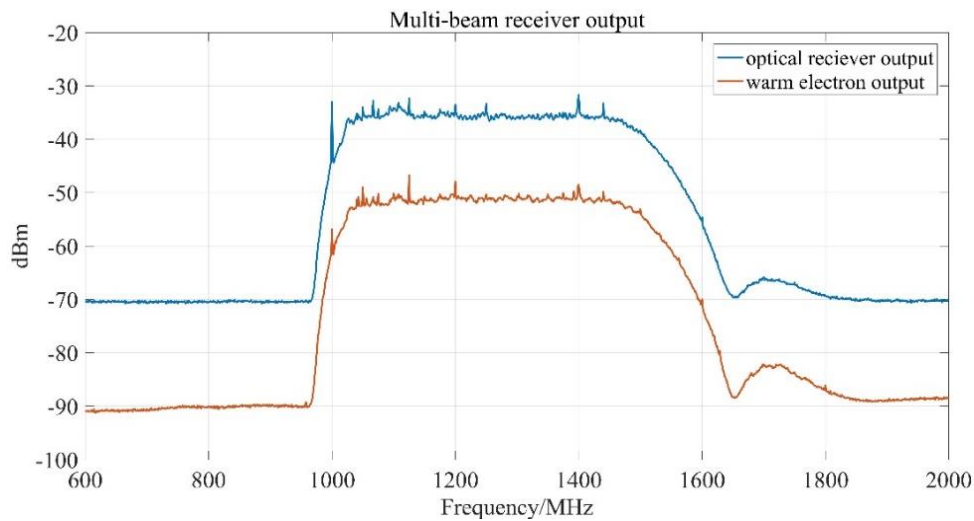

Fig. 4.5 Frequency coverage of the 19-beam 1.05-1.45 GHz receiver.

**2) Digital backends**

In the field of radio astronomy, digital backends have nearly replaced analog backends. Owing to the advancements in the semiconductor and electronic technology, the performance of FAST digital backends has been improved in comparison with the initial design. Meanwhile, to serve current scientific goals, new observation modes

**33** / **49**

have been added and the overall amount of data has increased in comparison with earlier estimates. Herein, we give a brief introduction of FAST backends. The design and implementation details will be presented in a future paper[1].

FAST backends can be classified as the pulsar de-dispersion backend, spectral-line backend, very-long-baseline interferometry (VLBI) data recording backend, search for extraterrestrial intelligence (SETI) backend, and baseband recording backend, according to their functions. According to the corresponding receivers, they can be classified as either multi- or single-beam backend. In most case, both classifications are applicable. For example, we have single- and multi-beam pulsar backend with different designs.

The pulsar incoherent backend uses a polyphase filter bank for channelization and dispersion delay compensation; the spectral-line backend of the multi-beam also collects spectrum data in the observation bandwidth by the polyphase filter bank, and then a spectral-line analysis can be conducted; the VLBI backend uses digital baseband converters and a data recording backend based on a hard disk; the SETI backend performs a spectrum analysis of 500-MHz bandwidth data with a frequency resolution of approximately 1 Hz; the baseband recording backend can record wideband signals, and the powerful computing power can perform flexible data analysis in various ways, such as coherent de-dispersion processing of pulsar observation data and strong time-varying signals, thereby recovering pulses signals with high time resolution.

Most of the first-step processing of FAST digital backends use the reconfigurable open-architecture computing hardware–version 2 (ROACH2)[2] board developed by the Collaboration for Astronomy Signal Processing and Electronics Research (CASPER) (https://casper.ssl.berkeley.edu/wiki/KatADC) at the University of California, Berkeley. The second step uses graphics processing unit(GPU) clusters to process data transmitted through an Ethernet. Most of the components are mature commercial products. The time system provides a reference 10-MHz signal from a hydrogen master clock. The frequency synthesizer uses this 10-MHz signal as a reference to provide clocks of different frequency requirements for sampling, thereby ensuring frequency accuracy. The pulse-per-second (1 PPS) signal of the hydrogen clock continuously monitors and compares the 1-PPS signal from the global positioning system (GPS) to ensure the traceability of time.

---

[1] Zhu, Y. et al., FAST backends, in preparation.
[2] Zhy, Y., private communication.

Since the start of FAST commissioning, the single-beam backend has been tested in ultra-wideband receivers for observation, and dozens of pulsars have been discovered. Since July, 2018, the multi-beam backend has been working with the 19-beam 1.05-1.45 GHz receiver, which has multiple commensal observation modes of spectral lines, pulsars, FRB, and SETI. Currently, the pulsar observation mode and spectral-line observation mode have been commissioned to work.

Fig. 4.6 shows the 19-beam 1.05-1.45 GHz backend, which is our most sophisticated backend. The entire backend comprises 12 × ROACH2, ten of which are online for most of the time, whereas two others are used for backup and new mode development. Multiple observation modes can be used simultaneously using Ethernet multicast technology, achieving the first step in the simultaneous observation of multiple scientific objectives. The pulsar, narrowband spectral line, wideband spectral line, FRB, and SETI modes can now work simultaneously.

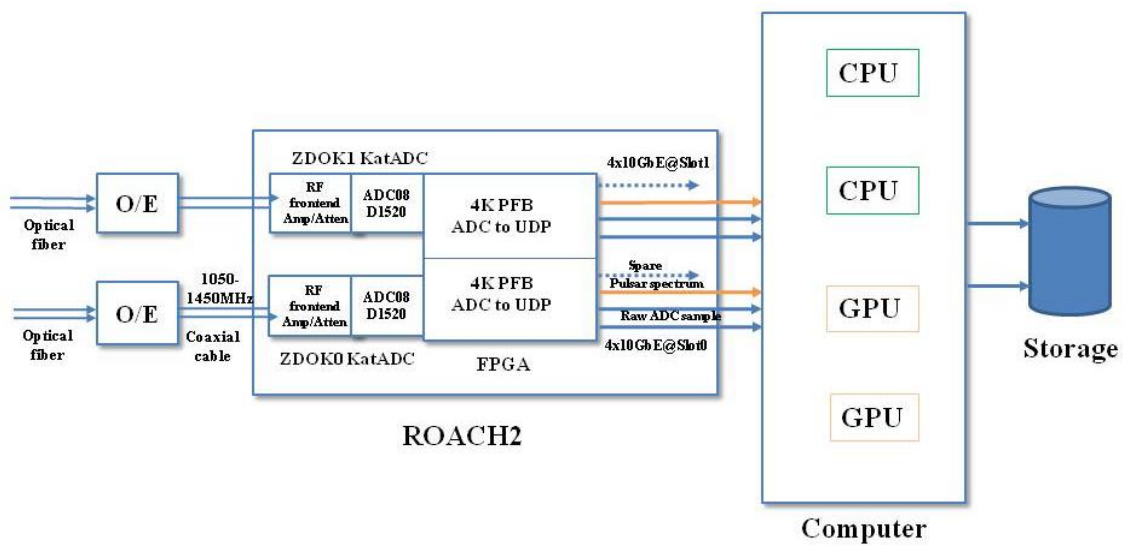

Fig. 4.6 The 19-beam backend diagram, updated from (https://casper.berkeley.edu/wiki/ROACH-2

_Revision_2). Here, of all the ROACH2 boards, only one is shown. Analog signals from the 19-beam receiver are transferred through fibers. After optical/electronic (O/E) conversion, the signal becomes a raw voltage signal. The raw voltage is sampled by analog-to-digital converter board KatADC (https://casper.berkeley.edu/). The digital signals are processed on the FPGA through a 4k-channel PFB. 1k- and 2k-channel modes are also available. Channelized data and raw sampling data are embedded in UDP packets and sent to the computer cluster for further processing. Generated FITS files are temporally stored on network-attached storage and then transferred to Guiyang for scientific processing.

### 4.2 Flux calibration and sensitivity test

The telescope system measures the power value of the instrument. To convert it to astronomically observable parameters, the brightness temperature (unit, K) and measured power value need to be calibrated so that the results can be compared with data obtained from other radio telescopes. Flux calibration is achieved by injecting a signal from a noise diode, the intensity of which has been calibrated in advance. The period of the noise diode provided by the FAST's 19-beam, 1.05–1.45 GHz receiver can be adjusted separately to synchronize with the sampling rate of the backend data. A high-intensity noise signal (~10 K) is generally used to calibrate the signal intensity. The variation of this noise intensity as a function of frequency is shown in Fig. 4.7.

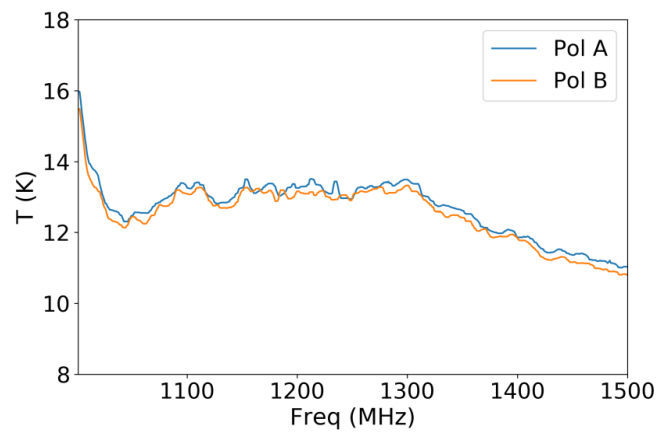

Fig. 4.7 Temperature variation of the high-intensity noise diode measured by beam 1 of the 19-beam, 1.05–1.45 GHz receiver; the blue and orange lines represent the measurements of two polarizations.

Telescope efficiency and sensitivity are generally measured by observing a radio source with a known flux density (e.g., 3C286 with steady flux density, Perley & Butler, 2013) [21]. Information on telescope efficiency and sensitivity can be obtained by comparing the measured radio power with the perfect paraboloid power.

The sensitivity of the FAST is evaluated using the direct ratio $R$ between the effective area $A_{eff}$ and system temperature $T_{sys}$. $R$ is inversely proportional to the system equivalent flux density (SEFD), which is generally used in astronomy. The larger the value of $R$ is, the higher the sensitivity will be:

$$R = \frac{1}{SEFD} = \frac{A_{eff}}{T_{sys}}. \tag{15}$$

$R$ depends on the telescope efficiency and system temperature. Owing to the fact that the FAST uses different reflector panels while pointing to different sky regions, the illumination area of the feed cabin varies. It is expected that $R$ will change along with the ZA.

During telescope commissioning, the flux calibrator, 3C286, was observed at different ZA in the drifting scan mode to derive the curve of $R$ as a function of ZA. As shown in Fig. 4.8, when ZA < 20.4°, the value of $R$ remains at ~2,400 m$^2$/K. The value of $R$ gradually decreases when ZA is greater than 20.4° and reaches about 2,000 m$^2$/K at ZA of 26.4°. When ZA is greater than 26.4°, $R$ decreases more rapidly with an increase in ZA due to loss in reflector panels and the influence of the feed illumination region beyond the paraboloid. The value of $R$ at ZA = 40° is about half than that of when ZA < 26.4°.

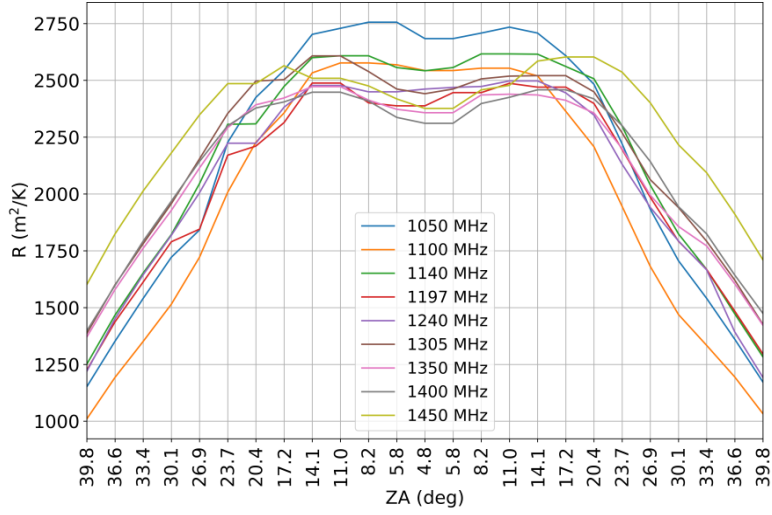

Fig. 4.8 *R*-value with different frequencies and ZAs when drifting 3C286.

The system temperature $T_{sys}$ can be calibrated by injection of the noise diode signal, thus $\eta$ can be measured independently. As shown in Figs. 4.9 and 4.10, when ZA varies in the range from 20.4° to 26.4°, $\eta$ stays constant at ~0.63. This means that the decrease in $R$ in this ZA range is due only to the increasing system temperature $T_{sys}$. At a ZA value of 40°, $\eta$ is 0.45, which decreases by only about 1/3 in comparison to that at ZA of 26.4°.

The gain of a radio telescope toward a point source in a single polarization, $G$, is expressed as

$$G = \eta G_0, \qquad (16)$$

where $G_0$ is the gain in single polarization without any loss. For the FAST, $G_0 =$ 25.2 K/Jy. Adopting the measured efficiency of 63%, $G$ of the FAST is 16.1 K/Jy. This value decreases to 11.3 K/Jy when ZA is ~40°.

The system temperature of the L-band of the Arecibo telescope in Puerto Rico is about 25–30 K and the gain $G$ is 9–11 K/Jy (http://www.naic.edu/~astro/RXstatus/ Lwide/Lwide.shtml). The performance of the FAST is expected to be better than that of the Arecibo telescope, even if the FAST loses some of its panels and the system temperature increases at ZA of 40°.

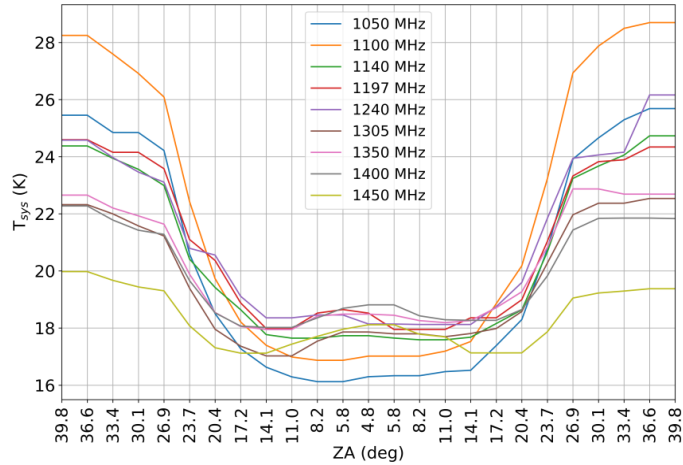

Fig. 4.9 System temperature as a function of frequency and ZA when drifting across 3C286.

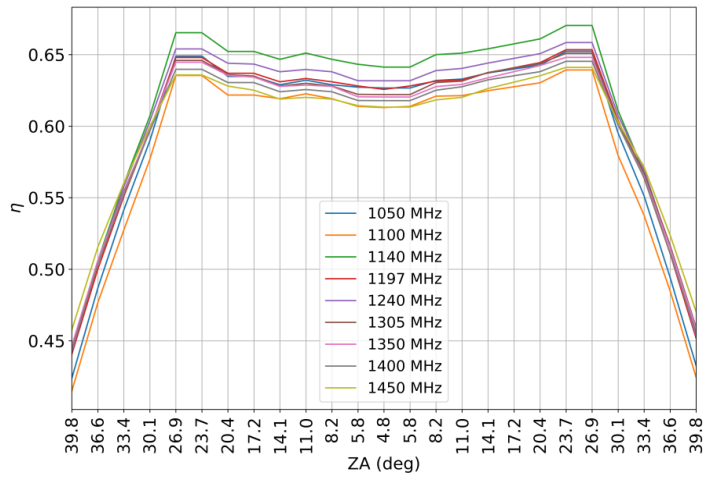

Fig. 4.10 Telescope efficiency as a function of frequency and ZA when drifting across 3C286.

As shown in Figs. 4.11–4.13, the telescope's tracking property was obtained by tracking 3C286 and its offset with 8'. The sensitivity, system temperature, and telescope efficiency in the tracking mode have a similar tendency as that under drift scanning.

In the future, the FAST project will further reduce the system temperature and construct a compensation model of a reflector surface to improve the accuracy of the active reflector system and the efficiency of the telescope during tracking.

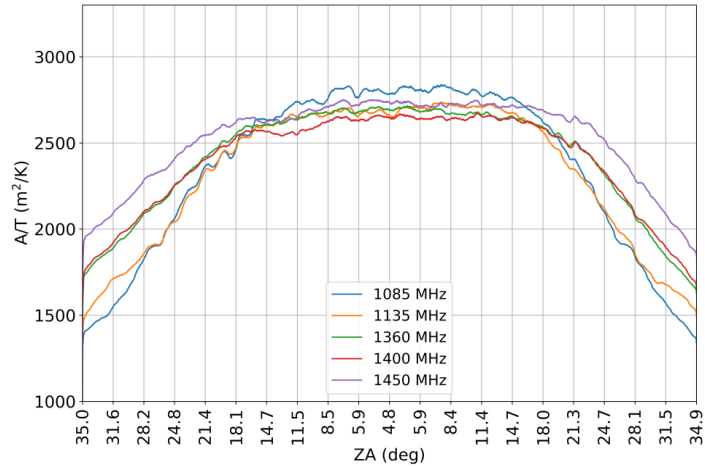

Fig. 4.11 *R*-value as a function of frequency and ZA when tracking 3C286 and the OFF position of 3C286.

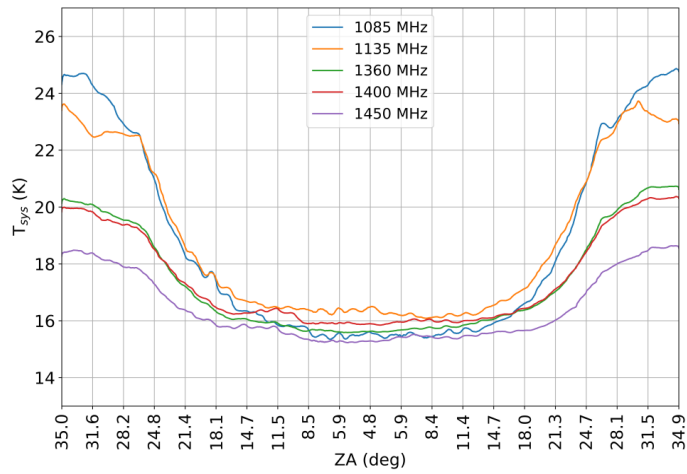

Fig. 4.12 System temperature as a function of frequency and ZA when tracking 3C286 and the OFF position of 3C286.

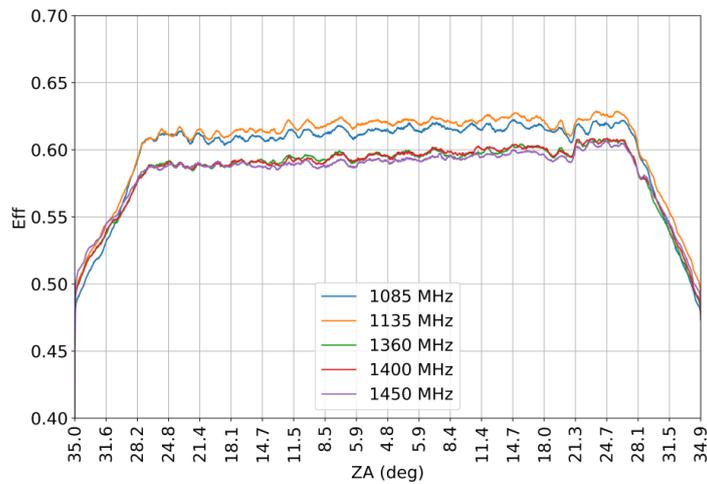

Fig. 4.13 Telescope efficiency as a function of frequency and ZA when tracking 3C286 and the OFF position of 3C286.

## 4.3 Beam shape test

An example of the beam pattern of the 19-beam receiver is presented in Fig. 4.14. The beam width as a function of frequency is shown in Fig. 4.15.

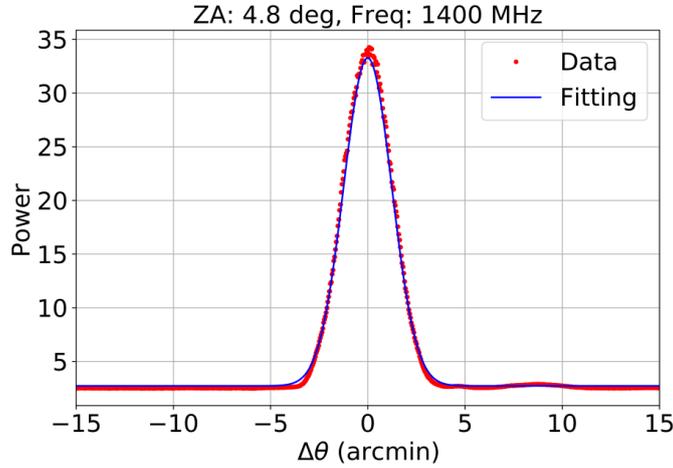

Fig. 4.14 Beam pattern of beam 1 at 1,400 MHz. The observation was taken toward 3C286 at a ZA of 4.8 deg. Data and fitting are shown by the red dots and blue solid line, respectively.

In Fig. 4.15, the observation was taken at a ZA of 4.8°. It shows the values for observation (solid lines), theoretical cosine-tapered illumination (dashed brown line), and theoretical uniform illumination (dashed pink line). These values were obtained by drift scanning across point calibrators. Theoretically, for a radio telescope with a diameter of D, full width at half maximum (FWHM) of the telescope beam at a wavelength $\lambda$ is $\theta_{\text{FWHM}} = 1.02 \frac{\lambda}{D}$ for a uniformly illuminated aperture and $\theta_{\text{FWHM}} = 1.22 \frac{\lambda}{D}$ for cosine-tapered illumination. It is obvious that the obtained beam behavior is among the uniform and cosine-tapered illumination.

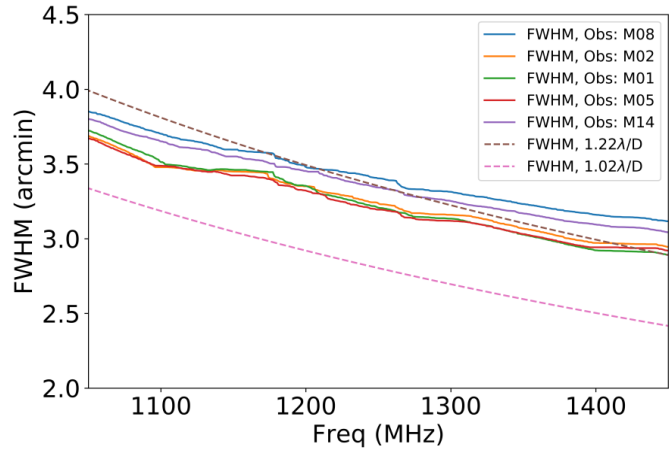

Fig. 4.15 Beam size as a function of frequency for beams 8, 2, 1, 5, and 14 of the 19-beam receiver.

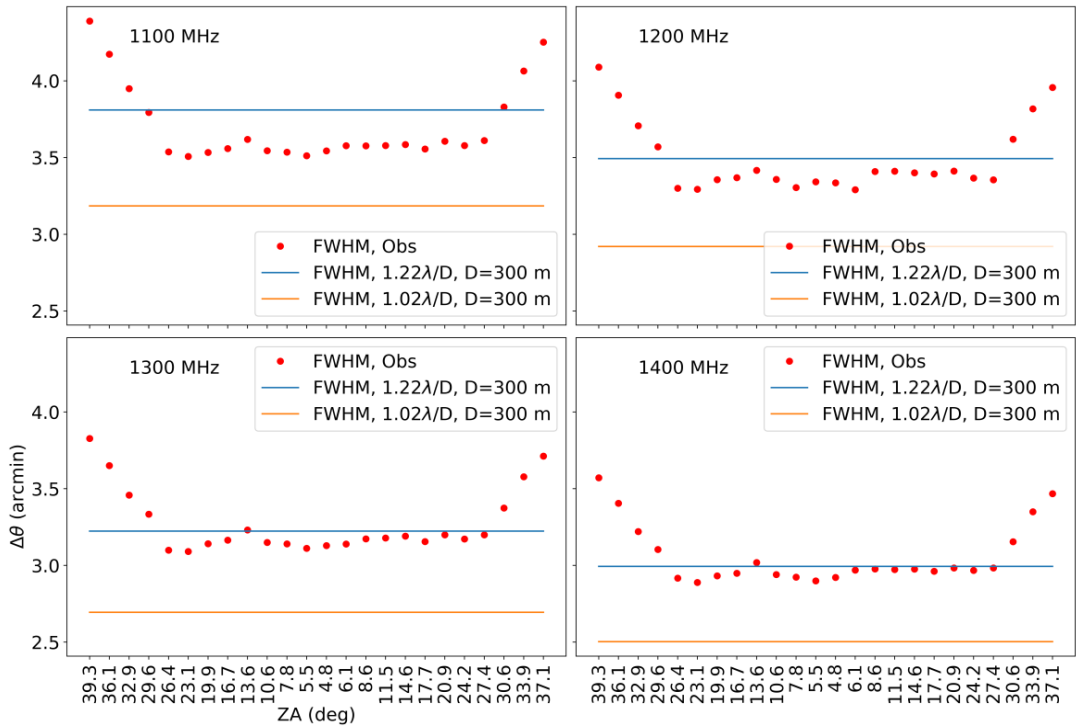

Fig. 4.16 Beam size as a function of ZA at four frequencies (1,100, 1,200, 1,300, and 1,400 MHz) for beam 1 of the 19-beam receiver.

As shown in Fig. 4.16, the beam size as a function ZA is presented. In this figure, the values for the observation (red dots), theoretical cosine-tapered illumination (solid blue line), and theoretical uniform illumination (solid orange line) are shown. A diameter of 300 m is assumed for the theoretical calculation. The beam size would obviously increase when ZA is greater than 26.4°. This is expected because the FAST would lose some of its panels when ZA > 26.4° and its effective surface area is small in comparison with a full paraboloid with a D of 300 m. It reaches a value of 3.5' at ZA = 40°, which corresponds to an effective D of 255 m.

### 4.4 Pointing calibration

The pointing of the FAST may deviate from the theoretical pointing. The pointing calibration aims to analyze the pointing deviation of the telescope by observing the calibration source. The pointing deviation can be used to rectify the pointing of the telescope and guide the observation data error analysis.

The calibration method of the FAST is to drift scan or basketweave scan several calibration sources, such as stable 3C sources (including 3C84 and 3C286), which are selected according to the declination distribution in the sky within the field of view of FAST. For a drifting scan, selecting the declination position of the source, the source is drift scanned by adding or subtracting 1.1' and 2.2' of the declination position; for basketweave scanning, the feedback source would grid scan (the representative scanning track is shown in the bottom left panel of Fig. 4.17) the square region centered by the source with a side length of 15' and scanning path separation of 1'. All the observation data are recorded by the pulsar backend. The noise diode is turned on during recording, and the switching frequency is exactly four times the sub-integration time of the recorded data. Finally, in the process of drift scanning, the real-time measurement system gives the pointing of the phase center of the feed.

After obtaining the relevant measurement data, the corresponding data processing is needed. The processing flow is as follows:

1) Read five groups of data of drift or basketweave scanning from the pulsar backend, judge and eliminate the noise according to the band pass condition, and then superimpose each group of data onto nine narrowband data with a bandwidth of 50 MHz in the range 1,025–1,475 MHz.

2) Calibrate the data from every four sub-integration obtained in the previous step by noise tube flow to obtain the relative flow series of the respective observations.

3) Read the start time of the data recorded at the pulsar backend. Considering the correction for leap seconds, the time series of the observations can be calculated from the sub-integration time, which corresponds to the flow series obtained in the previous step.

4) Read the three-dimensional coordinates of the phase center of the feed and measurement time from the measurement system and correct the leap seconds; an interpolation method is used to obtain the three-dimensional coordinates of the feed phase center corresponding to each flow data point.

5) Calculate the right ascension and declination using the three-dimensional coordinates obtained by the interpolation and the time corresponding to each flow data point. In this process, atmospheric correction is considered.

6) Perform a two-dimensional Gaussian fitting (no cross term) of the corresponding flow using the right ascension and declination of each data point to obtain the center position of the Gaussian peak corresponding to each frequency and the fitting error, which is compared with the theoretical position of the source.

Fig. 4.17 shows the results of the pointing test after the drift scanning on September 10, 2018 (top panels) and basketweave scanning on November 2, 2018 (bottom panels). The left panels show the scanning track of each observation. The center panels show Gaussian smoothing maps based on the narrowband flow of various frequencies on five scanning lines. In each figure, the black star represents the theoretical source position and the red point represents the position of the phase center of the feed by two-dimensional Gaussian fitting of the scanning lines. The right panels show comparisons of the fitting position with the theoretical position of the phase center of the feed. The red point is the theoretical position of the source. It can be seen that there is a certain difference among the phase center positions of various frequencies.

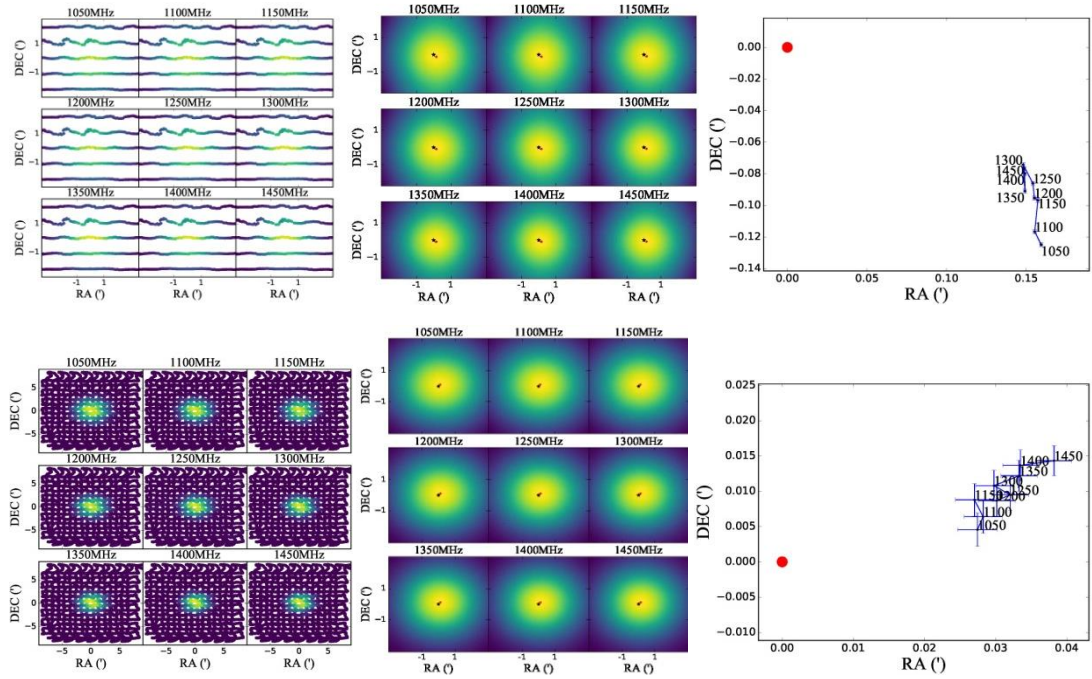

Fig. 4.17 Observation results of the pointing calibration.

According to the scanning observation results for September, it can be considered that the pointing deviation of the telescope is stable and that the deviation from the source is less than 0.2'. Table 4.1 shows the calibration results of several representative scans.

Table 4.1 The calibration results of several representative scans.

| Scan Mode | Source Name | Date | Source Direction | | Pointing Deviation | |
|---|---|---|---|---|---|---|
| | | | ZA (°) | AZ (°) | RA (") | DEC (") |
| **Drift Scanning** | 3C286 | 2018-09-10 | 10.4 | −53.2 | −7.2 | −4.8 |
| | 3C196 | 2018-09-19 | 24.0 | 15.5 | 1.9 | 1.2 |
| | 3C295 | 2018-11-01 | 27.5 | −11.4 | 16.0 | 8.5 |
| | 3C286 | 2018-11-05 | 10.3 | 53.0 | 8.8 | −1.0 |
| **Basketweave Scanning** | 3C295 | 2018-10-30 | 27.1 | −9.1 | 16.5 | −5.6 |
| | 3C286 | 2018-11-02 | 9.2 | 44.6 | 1.9 | 0.6 |
| | 3C196 | 2018-11-03 | 24.1 | 16.8 | 7.5 | −2.0 |

It should be noted that 3C295 has exceeded the ZA of 26.4°. At this time, the receiver adopts the backward illumination mode [22] and its beam shape and pointing performance need to be further tested and analyzed.

## 5. Future Prospects of the FAST Telescope

The FAST telescope is in the process of national acceptance preparations, although it is still striving to improve its capabilities and expand its application prospects:

### 5.1 Further calibration of the FAST

Considerable pointing calibration work still needs to be implemented. The calibration of the baseline and testing of the beam shape are important for spectral-line observation. Especially for large ZA observation, when the receiver adopts the backward illumination, the beam shape and pointing of the telescope need to be further tested and analyzed. Meanwhile, the calibration of the timing system is the foundation of timing observation and VLBI observation. To improve the performance of the FAST, a precise pointing model of the telescope is needed to further improve its pointing accuracy.

### 5.2 Fusion measurement method of total station and inertial navigation

Currently, the measurement equipment of the feed support system is the total station, which has the advantage of high distance measurement accuracy. However, the total station is also vulnerable to atmospheric environment impact, which will result in measurement accuracy loss and low dynamic measurement frequency. Considering other measurement methods, inertial navigation measurement has the characteristics of high-dynamic measurement frequency and independence of external information, which can work in all-weather conditions but there is a zero-drift problem. The global navigation satellite system (GNSS) is not affected by the meteorological environment and has all-weather advantages but the measurement accuracy is low [23].

Therefore, a single-measurement method cannot meet the feed support measurement requirement. It is therefore necessary to develop a fusion measurement system of the feed support, which is combined with the advantages of the above three measurement methods.

Currently, the FAST project plans to purchase two sets of inertial navigation components and develop a total station, inertial navigation, and GNSS data fusion processing algorithm, which is used by the fusion measurement system to provide high-precision, high-dynamic, and all-weather measurement data for the feed support system.

### 5.3 Application prospects of the ultra-wideband receiver

Currently, owing to a low-frequency coverage of the ultra-wideband receiver, as low as 270 MHz, the size of the feed horn is huge at 1.4 $m^2$ × 1.2 m; therefore, it cannot be cooled by the Dewar but works at normal temperature. At this temperature, the noise introduced from the feed horn and coaxial cable at its backend is more than 10 K. In the future, if the frequency coverage of the ultra-wideband receiver is properly increased to a high-frequency band as a whole, the size of the feed horn will be reduced accordingly until it can be cooled in the Dewar. This will greatly reduce the noise temperature of the receivers and improve the sensitivity of the telescope. This is a choice after balancing the scientific observation requirement and receiver manufacturing process.

**Acknowledgment**

This paper was financially supported by the National Natural Science Foundation of China (Nos. 11673039, 11573044, 11673002, 11803051, 11503048, 11203048); the Youth Innovation Promotion Association CAS; the Open Project Program of the Key Laboratory of FAST, NAOC, Chinese Academy of Sciences; the National Key R&D Program of China (2017YFA0402600) and CAS "Light of West China" Program; the Young Researcher Grant of National Astronomical Observatories, Chinese Academy of Sciences. The FAST FELLOWSHIP is supported by Special Funding for Advanced Users, budgeted and administrated by Center for Astronomical Mega-Science,

Chinese Academy of Sciences (CAMS).